\pdfoutput=1
\documentclass[10pt,oneside,reqno]{amsart}
\usepackage{amsfonts}
\usepackage{amsthm}
\usepackage{amsmath}
\usepackage{amscd}
\usepackage[latin2]{inputenc}
\usepackage{t1enc}
\usepackage[mathscr]{eucal}
\usepackage{indentfirst}
\usepackage{graphicx}
\usepackage{graphics}
\usepackage{rotating}
\usepackage{pdflscape}
\usepackage{pict2e}
\usepackage{epic}
\numberwithin{equation}{section}
\usepackage[margin=1in]{geometry}
\usepackage{epstopdf}
\usepackage{enumerate}
\usepackage[citestyle=numeric, bibstyle=numeric, backend=biber]{biblatex}
\usepackage{hyperref}

 \usepackage{booktabs}

\usepackage{pgfplots}
\pgfplotsset{compat=newest}
\usepgfplotslibrary{groupplots}
\usepgfplotslibrary{dateplot}
\usepackage{tikzscale}

\theoremstyle{definition}
\newtheorem{Th}{Theorem}[section]
\newtheorem*{Th*}{Theorem}

\theoremstyle{definition}
\newtheorem{Def}[Th]{Definition}
\newtheorem{Conj}[Th]{Conjecture}
\newtheorem*{Conj*}{Conjecture}
\newtheorem{Rem}[Th]{Remark}
\newtheorem{?}[Th]{Problem}
\newtheorem{Ex}[Th]{Example}

\newcommand{\R}{\mathbb{R}}

\usepackage{algorithm}
\usepackage{algpseudocode}

\algnewcommand{\IfThenElse}[3]{
  \State \algorithmicif\ #1\ \algorithmicthen\ #2\ \algorithmicelse\ #3}
  
\algnewcommand{\IfThen}[2]{
  \State \algorithmicif\ #1\ \algorithmicthen\ #2}

\usepackage{etoolbox}

\newcommand{\algstrut}[1][\algruledefaultfactor]{\vrule width 0pt
depth .25\baselineskip height #1\baselineskip\relax}
\newcommand*{\algrule}[1][\algorithmicindent]{\hspace*{.5em}\vrule\algstrut
\hspace*{\dimexpr#1-.5em}}

\makeatletter

\newcount\ALG@printindent@tempcnta
\def\ALG@printindent{%
    \ifnum \theALG@nested>0
    \ifx\ALG@text\ALG@x@notext
    \else
    \unskip
    \ALG@printindent@tempcnta=1
    \loop
    \algrule[\csname ALG@ind@\the\ALG@printindent@tempcnta\endcsname]%
    \advance \ALG@printindent@tempcnta 1
    \ifnum \ALG@printindent@tempcnta<\numexpr\theALG@nested+1\relax
    \repeat
    \fi
    \fi
}%

\usepackage{xcolor}
\definecolor{pink}{HTML}{F2059F}
\definecolor{green}{HTML}{04D94F}
\definecolor{orange}{HTML}{F28705}
\definecolor{maroon}{HTML}{BF0A3A}
\definecolor{blue}{HTML}{0D7DFF}
\definecolor{red}{HTML}{E85400}

\usepackage{tikz}
\usetikzlibrary{arrows.meta}
\usetikzlibrary{arrows,backgrounds,calc,trees}

\pgfdeclarelayer{background}
\pgfsetlayers{background,main}

\newcommand{\convexpath}[2]{
  [   
  create hullcoords/.code={
    \global\edef\namelist{#1}
    \foreach [count=\counter] \nodename in \namelist {
      \global\edef\numberofnodes{\counter}
      \coordinate (hullcoord\counter) at (\nodename);
    }
    \coordinate (hullcoord0) at (hullcoord\numberofnodes);
    \pgfmathtruncatemacro\lastnumber{\numberofnodes+1}
    \coordinate (hullcoord\lastnumber) at (hullcoord1);
  },
  create hullcoords
  ]
  ($(hullcoord1)!#2!-90:(hullcoord0)$)
  \foreach [
  evaluate=\currentnode as \previousnode using \currentnode-1,
  evaluate=\currentnode as \nextnode using \currentnode+1
  ] \currentnode in {1,...,\numberofnodes} {
    let \p1 = ($(hullcoord\currentnode) - (hullcoord\previousnode)$),
    \n1 = {atan2(\y1,\x1) + 90},
    \p2 = ($(hullcoord\nextnode) - (hullcoord\currentnode)$),
    \n2 = {atan2(\y2,\x2) + 90},
    \n{delta} = {mod(\n2-\n1,360) - 360}
    in 
    {arc [start angle=\n1, delta angle=\n{delta}, radius=#2]}
    -- ($(hullcoord\nextnode)!#2!-90:(hullcoord\currentnode)$) 
  }
}

\DeclareSymbolFont{extraup}{U}{zavm}{m}{n}
\DeclareMathSymbol{\varheart}{\mathalpha}{extraup}{86}
\DeclareMathSymbol{\clubsuit}{\mathalpha}{extraup}{84}

\patchcmd{\ALG@doentity}{\noindent\hskip\ALG@tlm}{\ALG@printindent}{}{\errmessage{failed to patch}}

\AtBeginEnvironment{algorithmic}{\lineskip0pt}

\begin{document}

\title{Minority Voter Distributions and Partisan Gerrymandering}

\author[J. Chen]{Jiahua Chen}

\author[A. Manne]{Aneesha Manne}

\author[R. Mendum]{Rebecca Mendum}

\author[P. Sahoo]{Poonam Sahoo}

\author[A. Yang]{Alicia Yang}

\makeatletter
\let\@wraptoccontribs\wraptoccontribs
\makeatother

\contrib[Mentored by:]{Diana Davis}


\begin{abstract}
    Many people believe that it is disadvantageous for members aligning with a minority party to cluster in cities, as this makes it easier for the majority party to \emph{gerrymander} district boundaries to diminish the representation of the minority. We examine this effect by exhaustively computing the average representation for every possible $5\times 5$ grid of population placement and district boundaries. We show that, in fact, it is \emph{advantageous} for the minority to arrange themselves in clusters, as it is positively correlated with representation. We extend this result to more general cases by considering the dual graph of districts, and we also propose and analyze metaheuristic algorithms that allow us to find strong lower bounds for maximum expected representation. 
\end{abstract}

\maketitle

\section{Introduction and Motivation}
Elections are integral to the operation of electoral democracies. In the legislative branch of the United States government, \emph{legislative bodies} such as the House of Representatives, as well as state legislatures, all consist of groups of representatives from various \emph{electoral districts}. Plurality elections are used to determine individual representatives from each electoral district. These electoral districts are partitioned through a \emph{districting plan} that assigns each \emph{census block}, the smallest geographic unit of population, to a district. Geometrically, this is the partitioning of a set of census blocks into districting plans. In partisan democracies such as the United States, the political makeup of such legislative bodies heavily dictates the overall political stance, causing the entire assembly to vote in favor of a specific \emph{political party}. Therefore, opposing political parties, especially the incumbent party, with the power and authority to create districting plans, often \emph{gerrymander} the districts to create biased districting plans that favor themselves. 

This current system does not yield \emph{proportional representation}, which means that the percentage of voters for a given political party does not yield a commensurate number of seats. In the state of Massachusetts, for example, it has been rigorously proved that it would be nearly impossible to create a districting plan that deviates from the current 9 Democrat, 0 Republican plan, although Republican candidates often receive 30\% to 40\% of the total votes \cite{Duchin2018LocatingMassachusetts}. Gerrymandering exploits the fact that districting does not necessarily yield proportional representation and is used by incumbent political parties to maximize the number of seats they win. 

There are several types of gerrymandering: \emph{partisan gerrymandering}, which benefits one party at the expense of another; \emph{racial gerrymandering}, which dilutes the political representation of racial minorities; and \emph{incumbent gerrymandering}, which protects incumbents from both parties and creates \emph{safe seats}---districts in which the incumbent is guaranteed a victory. While our research is focused on partisan gerrymandering, it has implications that extend to other forms of gerrymandering of a similar nature (specifically, gerrymandering that diminishes some binary indicator, such as political affiliation). 

\begin{figure}[h]
    \centering
    \includegraphics[height=1.8in]{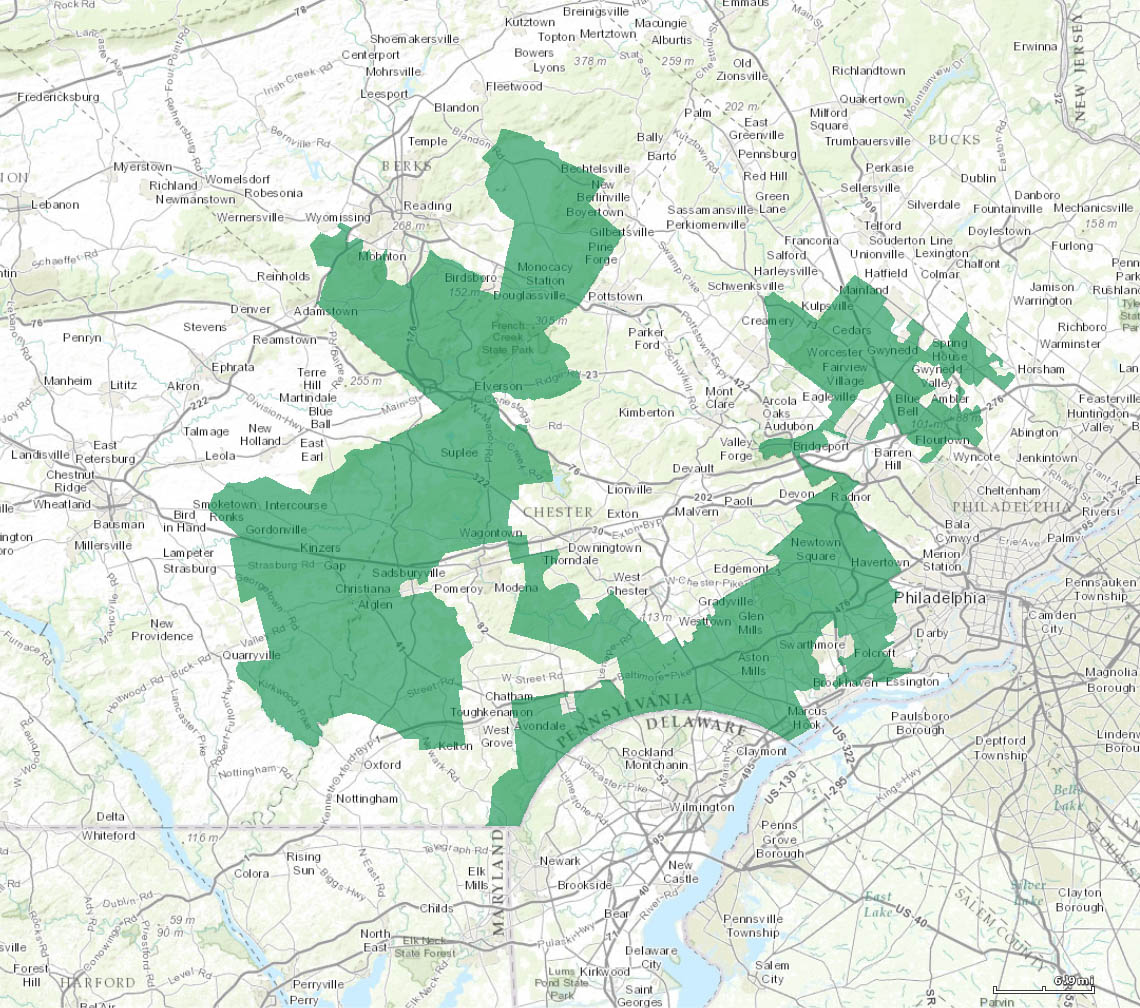}
    \includegraphics[height=1.8in]{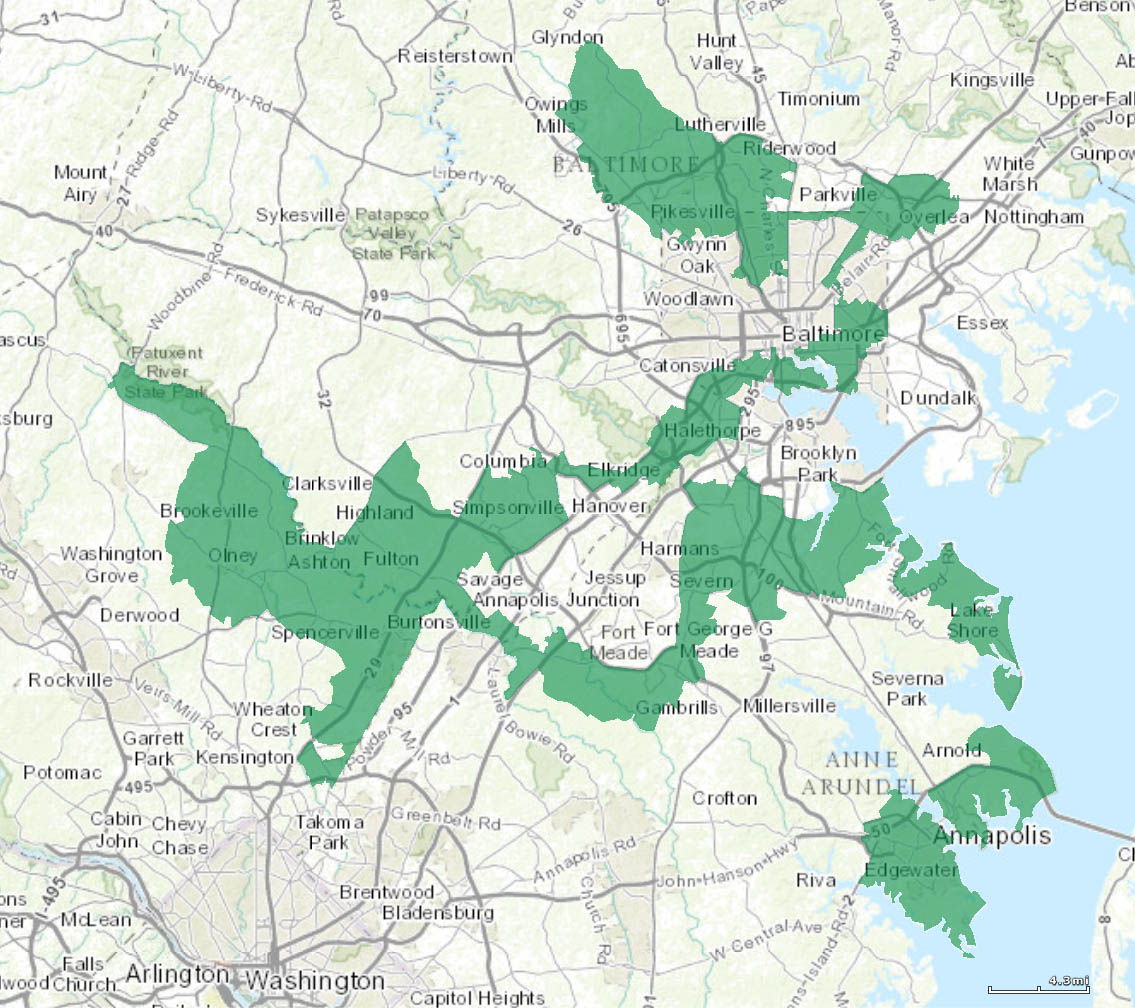}
    \includegraphics[height=1.8in]{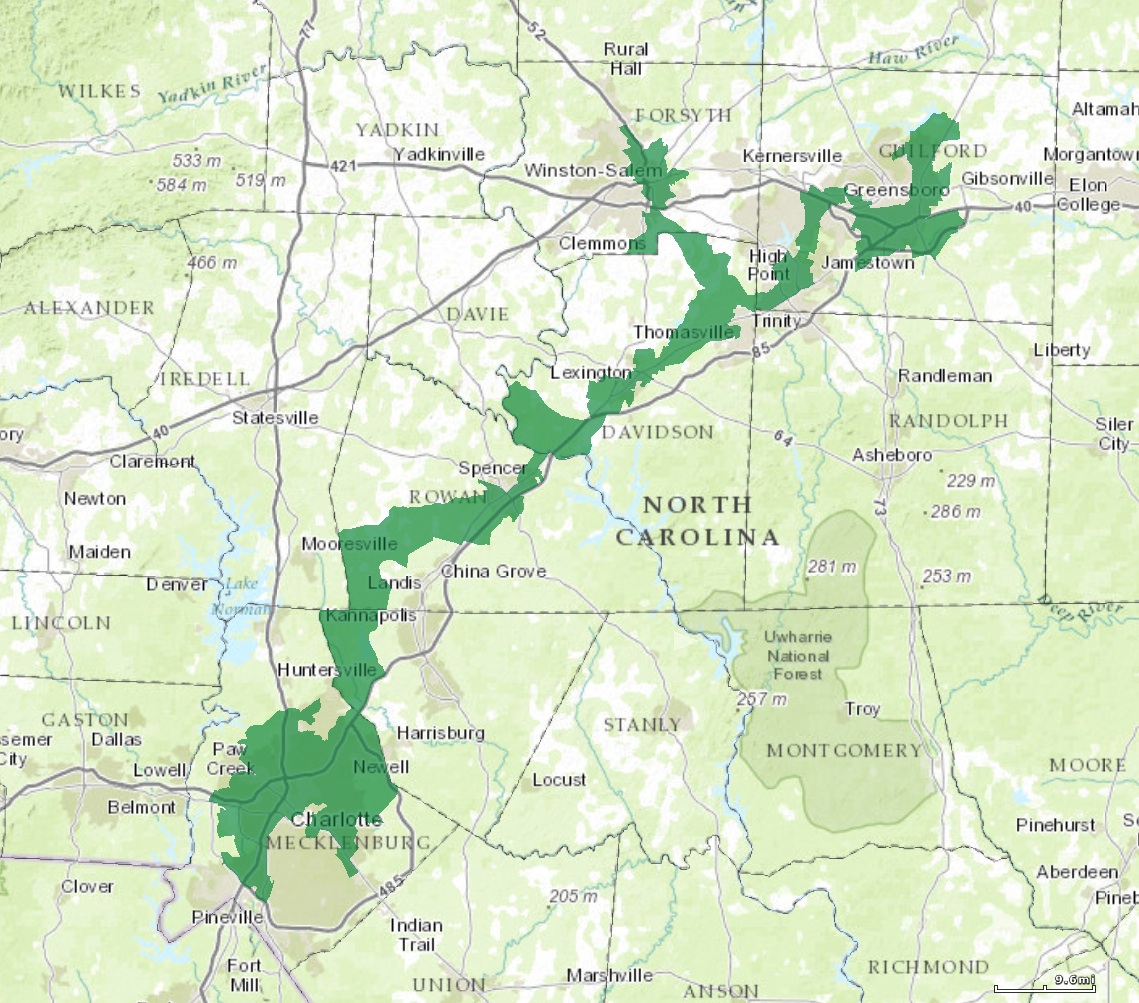}
    \caption{The congressional districts of PA-7 (prior), MD-3, and NC-12 (prior) demonstrate one of the geometrical consequences of gerrymandering---oddly shaped districts. The districting plans of Maryland and North Carolina have recently reached the Supreme Court. }
    \label{fig:districts}
\end{figure}

There are several methods for quantifying gerrymandering and `fairness' in elections. Recently, the Supreme Court of the United States saw the landmark case \emph{Rucho} v. \emph{Common Cause} (and \emph{Benisek} v. \emph{Lamone}) over the partisan gerrymander of North Carolina and Maryland's congressional districts. Although the Supreme Court came to the conclusion that ``partisan gerrymandering claims present political questions beyond the reach of the federal courts,'' leaving the case of partisan gerrymandering still hugely undecided, the court does however propose several methods for determining a fair districting plan \cite{2019RuchoAl.}. 

One method proposed by the Supreme Court (and commonly investigated in the mathematical community), is to determine the mean or median result out of a wide range of different districting plans. Because it is computationally difficult to enumerate through the space of all districting plans, such analysis is achieved by sampling the space of all districting plans using a stochastic Markov Chain Monte-Carlo method (also known as a Metropolis-Hastings algorithm). By constructing a \emph{metagraph} with nodes that are distinct districting plans and edges adjoining two similar districting plans, the algorithm makes a random walk on the graph to sample the space of districting plans \cite{Duchin2018GerrymanderingBaseline}. Each \emph{voter distribution} paired with a districting plan yields a different outcome (which is the number of seats for a given party). Using the algorithm, we can estimate the expected outcome for a given voter distribution over \textbf{all} the districting plans by evaluating the expected number of seats on a sample of all the districting plans. This is the measure of `fair' representation that courts and mathematicians propose---the mean result yielded by all the plans. 

This measure has significant implications. As the mean of the distributions of varying districting plans, it is an indicator for the overall distribution itself. A higher mean signifies that a districting plan yielding a high number of seats is more likely, while a lower mean signifies that a plan yielding a similar number of seats is more difficult to create and statistically unlikely. Thus, it is in the best interest of every political party to maximize this metric for themselves. In other words, every political party wants a voter distribution that yields the highest mean over all districting plans. This \emph{robust} voter distribution gives that party the best shot at gaining a higher amount of representation (seats) in the House or other legislative bodies. 

Our research is motivated by the following questions, which are closely related to this metric: 
\begin{enumerate}[1.]
\item \emph{How do certain attributes of a given voter distribution, if any, affect this measure? }
\item \emph{Are there specific ways of arranging a limited number of voters on a map that yields the best result for a certain political party? }
\item \emph{How can we compute the optimal voter distribution that yields the best result? }
\end{enumerate}

Specifically, we are interested in cases where the specific party is the minority party in a given state, as minority parties are often disadvantaged the most by malicious gerrymandering (such as the case in North Carolina in 2018). 

Chen and Rodden propose that one key characteristic affecting a particular voter distribution's expected representation is \emph{clustering}---a factor of human geography that affects the concentration of certain voters in cities or specific regions. By analyzing maps and election data, they come to the conclusion that Democrats disadvantage themselves by clustering in cities \cite{Chen2013UnintentionalLegislatures}. 

Whereas Chen and Rodden analyzed real-world data, we examine this effect on a theoretical level using a model that simulates the geographical landscape of a state: a $5\times 5$ grid population of voters. By exhaustively computing every possible combination of voter placement on this grid and determining the average representation (the above metric) for each of these voter placements, we are able to eliminate many confounding complexities that are associated with human geography in order to come to stronger conclusions regarding the correlation between clustering and expected representation. This is detailed in Section \ref{sec:grid}. 

We show that, in fact, it is advantageous for a minority party to arrange themselves in clusters, as it is positively correlated with expected representation. By representing an area of land as dual graphs, we are then able to extend this result to the general case, which we propose as a conjecture in Section \ref{sec:grid}. 

\begin{Conj*}
\label{th:clustering-relation}
	\textbf{(Clustering Relation).} For minority voter populations, there is a positive correlation between a measure of clustering and their expected representation. 
\end{Conj*}

Exploiting this relation, we are able to construct metaheuristic algorithms that allow us to optimize voter distributions to yield the maximum expected representation by using a greedy algorithm. We compare such an algorithm against other discrete metaheuristic optimization techniques in Section \ref{sec:metaheuristics}.

\section{Basic Definitions and Terminology}
\label{sec:definitions}

We first define some basic definitions and terminology that we use throughout the paper. 

\begin{Def}
We model population using a \emph{dual graph}, a planar, undirected graph $G = (V,E)$ consisting of $k$ vertices $V = \{b_1,b_2,\dots,b_k\}$, which we call blocks (since the vertices represent census blocks of equal population). If $b_i$ and $b_j$ are connected by an edge, they are \emph{adjacent} blocks. If $b_k$ is on the boundary of the unbounded face of the graph, we say $b_k$ is a \emph{border}. If two census blocks touch at a corner, they are not considered as sharing an edge, which ensures we get a planar graph. 
\begin{figure}[h]
    \centering
    \begin{tikzpicture}
\draw [fill=orange!80] (0, 0) rectangle (1, 2);
\draw [fill=green!30!blue!20] (0, 2) rectangle (1, 3);
\draw [fill=pink!70] (1, 2) rectangle (3, 3);
\draw [fill=green] (3, 2) rectangle (4, 3);
\draw [fill=maroon!80] (1, 0) rectangle (2, 1);
\draw [fill=yellow] (1, 1) rectangle (2, 2);
\draw [fill=blue!80] (2, 1) rectangle (3, 2);
\draw [fill=red!90] (2, 0) -- (4, 0) -- (4, 2) -- (3, 2) -- (3, 1) -- (2, 1) -- cycle;

\begin{scope}[every node/.style={circle,draw}, xshift=6cm]
    \node [fill=orange!80] (A) at (1/2, 1) {};
    \node [fill=green!30!blue!20] (B) at (1/2, 5/2) {};
    \node [fill=maroon!80] (C) at (3/2, 1/2) {};
    \node [fill=red!90] (D) at (7/2, 1/2) {};
    \node [fill=yellow] (E) at (3/2, 3/2) {};
    \node [fill=blue!80] (F) at (5/2, 3/2) {};
    \node [fill=green] (H) at (7/2, 5/2) {};
    \node [fill=pink!70] (G) at (2, 5/2) {};
    \draw (A) -- (B) -- (G) -- (H) -- (D) -- (C) -- (A) -- (E) -- (F) -- (G) -- (E) -- (C);
    \draw (F) -- (D);
\end{scope}
\end{tikzpicture}
    \caption{A region of 8 elementary census blocks (left) is represented as a dual graph (right) of 8 vertices (each representing an elementary census block), sharing an edge when two corresponding blocks are adjacent. }
    \label{fig:dual_graph}
\end{figure}
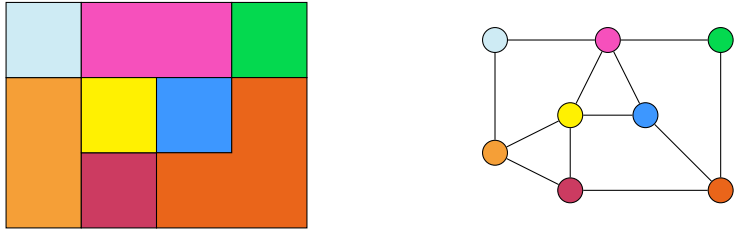
\end{Def}

For each dual graph, we can create \emph{districting plans} that assign each block to a unique district. Many of the `rules' regarding districting arise from the Voting Rights Act of 1965, which placed restrictions on districting in the United States \cite{Gavins2016Voting1965}. 

\begin{Def}
A \emph{districting plan} $\mathcal{D}$ partitions a dual graph $G$ into $n$ \emph{districts} $\{d_1,d_2,\dots,d_n\}.$ Then, a \emph{legal} districting plan is one that satisfies the following:

\begin{enumerate}[1.]
\item (Population) Each district $d$ contains the same number of blocks. Since blocks represent census blocks of equal population, our districts will also have equal population. The number of blocks is the \emph{size} $m$ of a district. 
\item (Contiguity) For all census blocks $b_x,b_y$ in a district $d_j$, $b_x$ and $b_y$ are connected by a series of edges with vertices all in $d_j$. In addition, if $b_{\alpha}$ is a block not in district $d_j$, there exists a series of edges connecting $b_{\alpha}$ to a border $b_{\lambda}$ in which none of the vertices pass through $d_j$. 
\end{enumerate}
The set of all districting plans is denoted by $\mathbb{D} = \{\mathcal{D}\mid \mathcal{D}\ \mathrm{legal}\}$.
\end{Def}
\begin{figure}[h]
    \centering
    \begin{tikzpicture}
        \begin{scope}[every node/.style={circle,draw}]
            \node [fill=orange!80] (A) at (1/2, 1) {};
            \node [fill=green!30!blue!20] (B) at (1/2, 5/2) {};
            \node [fill=maroon!80] (C) at (3/2, 1/2) {};
            \node [fill=red!90] (D) at (7/2, 1/2) {};
            \node [fill=yellow] (E) at (3/2, 3/2) {};
            \node [fill=blue!80] (F) at (5/2, 3/2) {};
            \node [fill=green] (H) at (7/2, 5/2) {};
            \node [fill=pink!70] (G) at (2, 5/2) {};
            \draw (A) -- (B) -- (G) -- (H) -- (D) -- (C) -- (A) -- (E) -- (F) -- (G) -- (E) -- (C);
            \draw (F) -- (D);
            \begin{pgfonlayer}{background}
            	\fill[red,opacity=0.3] \convexpath{A, B}{10pt};
            	\fill[blue,opacity=0.3] \convexpath{E, C}{10pt};
        		\fill[green,opacity=0.3] \convexpath{F, D}{10pt};
        		\fill[purple,opacity=0.3] \convexpath{H, G}{10pt};
            \end{pgfonlayer}
        \end{scope}
        \begin{scope}[every node/.style={circle,draw}, xshift = 5cm]
            \node [fill=orange!80] (A) at (1/2, 1) {};
            \node [fill=green!30!blue!20] (B) at (1/2, 5/2) {};
            \node [fill=maroon!80] (C) at (3/2, 1/2) {};
            \node [fill=red!90] (D) at (7/2, 1/2) {};
            \node [fill=yellow] (E) at (3/2, 3/2) {};
            \node [fill=blue!80] (F) at (5/2, 3/2) {};
            \node [fill=green] (H) at (7/2, 5/2) {};
            \node [fill=pink!70] (G) at (2, 5/2) {};
            \draw (A) -- (B) -- (G) -- (H) -- (D) -- (C) -- (A) -- (E) -- (F) -- (G) -- (E) -- (C);
            \draw (F) -- (D);
            \begin{pgfonlayer}{background}
            	\fill[red,opacity=0.3] \convexpath{B, G}{10pt};
            	\fill[blue,opacity=0.3] \convexpath{A, C}{10pt};
        		\fill[green,opacity=0.3] \convexpath{E, F}{10pt};
        		\fill[purple,opacity=0.3] \convexpath{H, D}{10pt};
            \end{pgfonlayer}
        \end{scope}
    \end{tikzpicture}
    \caption{Two \emph{legal} districting plans of $m=2$ for the dual graph in Figure \ref{fig:dual_graph}.}
    \label{fig:districting_plan}
\end{figure}
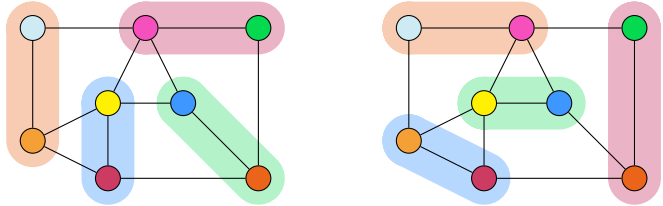

While the dual graph and districting plans are a representation of the geographic distribution of populations and districts, they do not contain information regarding the voter preferences of the people living in each individual block. 

\begin{Def}
\label{def:voter-distribution}
We define a \emph{discrete voter distribution} as a function $\Delta: V\to \{0,1\}$, mapping each vertex from the set of vertices of our dual graph to a binary measure $0$ or $1$. A measure of $1$ means that the block leans toward the given party, and a measure of $0$ means that the block leans away from the given party. Naturally, we can find the proportion of voters for a given party over the whole dual graph $G$: 
\begin{equation}
    \mathsf{Avg}_G\left(\Delta\right)=\frac{1}{k}\sum^k_{i=1}\Delta(b_i).
\end{equation}
Similarly, the proportion of voters for a given party in a given district $d_j$ is:
\begin{equation}
    \mathsf{Avg}_{d_j}\left(\Delta\right)=\frac{1}{m}\sum_{b_i\in d_j}\Delta(b_i);\quad m=\left|d_j\right|.
\end{equation}
In a bipartisanship, not leaning toward the party of interest implies a lean toward the opposing party. For the purposes of this paper, we let the party of interest (the `given party') be the {\color{red}$\bullet$} party. The set of all possible voter distributions is denoted by set $\{\Delta\}$. 
\begin{figure}[h]
    \centering
    \begin{tikzpicture}
\begin{scope}[every node/.style={circle,draw}]
    \node (A) at (1/2, 1) {\color{red}$\bullet$};
    \node (B) at (1/2, 5/2) {$\phantom{\bullet}$};
    \node (C) at (3/2, 1/2) {$\phantom{\bullet}$};
    \node (D) at (7/2, 1/2) {\color{red}$\bullet$};
    \node (E) at (3/2, 3/2) {$\phantom{\bullet}$};
    \node (F) at (5/2, 3/2) {\color{red}$\bullet$};
    \node (H) at (7/2, 5/2) {$\phantom{\bullet}$};
    \node (G) at (2, 5/2) {\color{red}$\bullet$};
    \draw (A) -- (B) -- (G) -- (H) -- (D) -- (C) -- (A) -- (E) -- (F) -- (G) -- (E) -- (C);
    \draw (F) -- (D);
\end{scope}
\end{tikzpicture}
    \caption{A discrete voter distribution for the dual graph in Figure \ref{fig:dual_graph}, where the party of interest is the {\color{red}$\bullet$} party, and the opposing party is an empty node.}
    \label{fig:voter_distribution}
\end{figure}
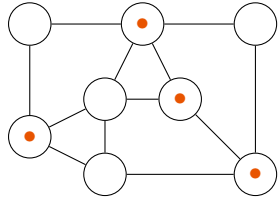
\end{Def}

For each voter distribution, we need a measure to quantify \emph{clustering} of a given party, which is an important indicator in our later analysis. 
\begin{Def}
\label{def:clustering}
\emph{Clustering} $\mathsf{Clus}:\{\Delta\}\to [0,1]$ of a voter distribution is the proportion of edges that join two vertices voting for the same party: 
\begin{equation}
    \mathsf{Clus}(\Delta)=\frac{\#\Big(\Delta(b_i) = \Delta(b_j)\Big)}{\#E}; \quad (b_i, b_j)\in E.
\end{equation}
We can also define a similar measure, \emph{partisan clustering} $\mathsf{ClusP}$ for, which comes with an additional rule to only count \textbf{directed} connections containing a block voting for our given party (using our definition above, means that $\Delta(b)=1$):
\begin{equation}
    \mathsf{ClusP}(\Delta)=\frac{\#\Big(\Delta(b_i) = \Delta(b_j)=1\Big)}{\#\{(b_i, b_j)\in E\mid  \Delta(b_i) = 1\}}.
\end{equation}
Here, $\#$ simply denotes the \emph{count of}.

To compute $\mathsf{Clus}$, we enumerate all the edges on the graph and divide the number of like edges by the total number of edges. Similarly, to compute $\mathsf{ClusP}$, we enumerate all directed edges that start at a {\color{red}$\bullet$} block and divide the number of edges connecting a {\color{red}$\bullet$} block to another {\color{red}$\bullet$} block over the total number of edges containing {\color{red}$\bullet$} blocks. 
\end{Def}

\begin{Ex}
For the voter distribution $\Delta$ in Figure \ref{fig:voter_distribution}, 
\begin{equation*}
	\mathsf{Clus}(\Delta) = \frac{3}{12} \quad\text{and}\quad \mathsf{ClusP}(\Delta) = \frac{4}{13}. 
\end{equation*}
Note that for $\mathsf{ClusP}$, we are double-counting each edge connecting a {\color{red}$\bullet$} to a {\color{red}$\bullet$} as we are concerned with directed edges. This emphasizes the clustering of the {\color{red}$\bullet$} party, which creates a more representative measure of clustering, especially when {\color{red}$\bullet$} is the minority party.  
\end{Ex}

We now want to be able to evaluate the \emph{representation} of a district to determine which party wins the seat of that district. 

\begin{Def}
The \emph{representation} of a district $d_i$ is the outcome of the plurality election held in the district. It is a function $R: \mathcal{D}\times \{\Delta\} \to \{0,0.5,1\}$ that takes a voter distribution and a given district in our districting plan and assigns the district a value of $0$, $0.5$, or $1$. $R$ evaluates to $0$ if the {\color{red}$\bullet$} party loses the seat in said district, $R$ evaluates to $1$ if the {\color{red}$\bullet$} party wins the seat in said district, and $R$ evaluates to $0.5$ if there is an equal number of blocks in the district voting for both parties.
\begin{equation}
    R_\Delta(d_i) = \begin{cases}
    0 & \text{if }\mathsf{Avg}_{d_j}(\Delta)< 0.5 \\
    0.5 & \text{if }\mathsf{Avg}_{d_j}(\Delta)= 0.5 \\
    1 & \text{if }\mathsf{Avg}_{d_j}(\Delta)> 0.5 
    \end{cases}.
\end{equation}
The \emph{total representation} $\mathsf{Rep}:\mathbb{D}\times \{\Delta\}\to [1, m]$ is the number of seats that a given party wins over \textbf{all} $m$ districts for a given voter distribution $\Delta$ and districting plan $\mathcal{D}$: 
\begin{equation}
    \mathsf{Rep}_\Delta(\mathcal{D}) = \sum_{d_i\in \mathcal{D}}R(d_i, \Delta).
\end{equation}
\end{Def}

We need a measure for each voter distribution $\Delta$ that does not depend on which districting plan we use. In other words, we need a measure that encapsulates the representation of a given voter distribution over \textbf{all} possible districting plans. We can achieve this by treating $\mathsf{Rep}_\Delta$ as a random variable. 

\begin{Def}
Another way we interpret $\mathsf{Rep}_\Delta$ is as a discrete random variable. Consider a sample space $\Omega = \mathbb{D}$ of all the districting plans where an event is selecting a particular districting plan $\mathcal{D}$ from our sample space uniformly. $\mathsf{Rep}_\Delta: \Omega \to \R$ is then our discrete random variable, which is the representation corresponding to a voter distribution. This allows us to do the following: 
\begin{enumerate}[1.]
\item We can create a probability density function for $\mathsf{Rep}_\Delta$, which shows how likely each level of representation is for our given party. 
\item We can compute the \emph{expected amount of representation} for our given party: 
\begin{equation}
    \label{eq:expected-representation}
    \mathbb{E}(\mathsf{Rep}_\Delta) = \frac{1}{|\mathbb{D}|}\sum_{\mathcal{D}\in\mathbb{D}}\mathsf{Rep}_\Delta (\mathcal{D}).
\end{equation}
\item We can compute the \emph{variance} of representation $\mathsf{Var}(\mathsf{Rep}_\Delta)$.
\item We can compute the \emph{minimum} and \emph{maximum} representation that could be won: $\mathsf{min}(\mathsf{Rep}_\Delta)$ and $\mathsf{max}(\mathsf{Rep}_\Delta)$. 
\end{enumerate}
\end{Def}

\begin{Rem}
The expected amount of representation in equation \ref{eq:expected-representation} becomes a key measure in our analysis as it provides a precise quantification of how well each districting plan $\Delta$ performs. However, $\mathbb{D}$ is usually very large, so it is computationally hard to compute this expected value. This also becomes a main focus and rate limiting factor in our optimization. 
\end{Rem}

\section{Grid}
\label{sec:grid}

In the United States, laws often restrict the process of redistricting to complying with county or town boundaries. Especially in the Midwest, land is often divided into grid-like \emph{precincts}, which are the fundamental units of voting districts. We can easily use a dual graph representing a square grid, which resembles the often square or rectangular regions of land. 

\begin{figure}[h]
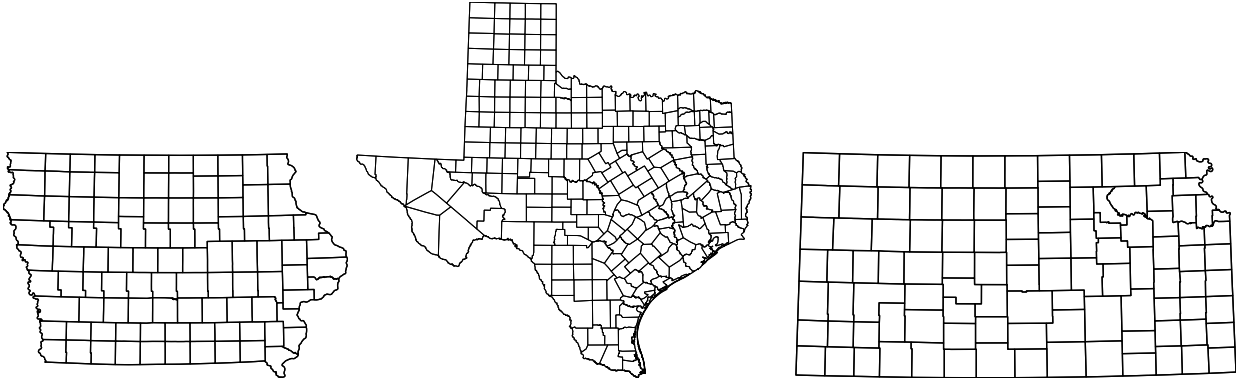

    \centering
    \includegraphics[height=3cm]{images/iowa.pdf}
    \includegraphics[height=5cm]{images/texas.pdf}
    \hspace{1em}
    \includegraphics[height=3cm]{images/kansas.pdf}
    \caption{Maps of the counties of Iowa, Texas, and Kansas. These states in the Midwest have counties that resemble a rectangular grid. }
    \label{fig:counties}
\end{figure}

While many states do not follow this geographical pattern, square grids are a computationally accessible model that gives a surprising amount of insight into the clustering problem. Additionally, the total number of possible legal districting plans, which we denote by $\mathbb{D}$, can be easily computed for small cases. 

Each square of the grid is a block of equal population. Following Definition \ref{def:voter-distribution}, we can also have voter distributions for our grids. 

\begin{figure}[h]
    \centering
    \begin{tikzpicture}[scale=.65]
\foreach \x in {0,...,5} {\draw [gray!85] (\x,0)--(\x,5);}
\foreach \y in {0,...,5} {\draw [gray!85] (0,\y)--(5,\y);}
\draw [line width=1.0] (0,0)--(5,0)--(5,5)--(0,5)--cycle;
\foreach \p/\q in {3/5, 4/5, 5/5, 1/4, 2/4, 3/4, 4/4, 1/3, 2/3, 1/2}  {\node at (\p-.5,\q-.5) {\color{red}$\bullet$};}
\end{tikzpicture}
    \caption{A $5\times 5$ \emph{square grid} with a voter distribution yielding $10$ {\color{red}$\bullet$} blocks. }
    \label{fig:square-grid}
\end{figure}
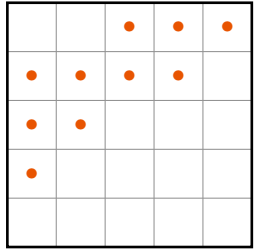

For a square grid, all legal districting plans are formed by \emph{polyominoes}, which are plane geometric figures formed by joining one or more equal squares edge to edge. The number of distinct polyomino partitions of square grids is well known for small dimensions, and the algorithms for finding such partitions are also optimized \cite{Harris2010CountingIlk}. For a $5\times 5$ grid of pentomino tilings (5 districts of 5 blocks each), there are a total of $4006$ possible districting plans.  

\begin{Rem}
Even with efficient algorithms to find pentomino tilings, this problem of tiling is still NP-complete. The rate at which the number of possible partitions increases exceeds $O(e^n)$. The known numbers of these tilings are listed in Table \ref{tab:polyomino-tilings}. 
\end{Rem}

\begin{table}[h]
    \centering
    \begin{tabular}{@{}rl@{}}
\toprule
$n$ & $n$-omino tilings        \\ \midrule
1    & 1 \\
2    & 2 \\
3    & 10  \\
4    & 117  \\
5    & 4006  \\
6    & 451206  \\
7    & 158753814  \\
8    & 187497290034  \\
9    & 706152947468301 \\ \bottomrule \\
\end{tabular}
    \caption{Number of $n$-omino tilings on a $n\times n$ square grid (\href{https://oeis.org/A172477}{OEIS: A172477)} \cite{Harris2010CountingIlk}. }
    \label{tab:polyomino-tilings}
\end{table}

\begin{figure}[h]
    \centering
    \begin{tikzpicture}[scale=.65]
    \begin{scope}
\foreach \x in {0,...,5} {\draw [gray!85] (\x,0)--(\x,5);}
\foreach \y in {0,...,5} {\draw [gray!85] (0,\y)--(5,\y);}
\draw [line width=1.0] (0,0)--(5,0)--(5,5)--(0,5)--cycle;
\foreach \p/\q in {3/5, 4/5, 5/5, 1/4, 2/4, 3/4, 4/4, 1/3, 2/3, 1/2}  {\node at (\p-.5,\q-.5) {\color{red}$\bullet$};}
\draw [line width=1.0, fill=red, fill opacity=.3] (0,0)--(3,0)--(3,2)--(1,2)--(1,1)--(0,1)--cycle;
\draw [line width=1.0, fill=purple, fill opacity=.3] (3,0)--(5,0)--(5,1)--(4,1)--(4,4)--(3,4)--cycle;
\draw [line width=1.0, fill=blue, fill opacity=.3] (0,1)--(1,1)--(1,2)--(2,2)--(2,4)--(0,4)--cycle;
\draw [line width=1.0, fill=green, fill opacity=.3] (0,4)--(2,4)--(2,2)--(3,2)--(3,5)--(0,5)--cycle;
\draw [line width=1.0, fill=pink, fill opacity=.3] (3,4)--(4,4)--(4,1)--(5,1)--(5,5)--(3,5)--cycle;
\end{scope}
\begin{scope}[xshift = 6cm]
\foreach \x in {0,...,5} {\draw [gray!85] (\x,0)--(\x,5);}
\foreach \y in {0,...,5} {\draw [gray!85] (0,\y)--(5,\y);}
\draw [line width=1.0] (0,0)--(5,0)--(5,5)--(0,5)--cycle;
\foreach \p/\q in {3/5, 4/5, 5/5, 1/4, 2/4, 3/4, 4/4, 1/3, 2/3, 1/2}  {\node at (\p-.5,\q-.5) {\color{red}$\bullet$};}
\draw [line width=1.0, fill=pink, fill opacity=.3] (0,0)--(4,0)--(4,2)--(3,2)--(3,1)--(0,1)--cycle;
\draw [line width=1.0, fill=purple, fill opacity=.3] (0,3)--(4,3)--(4,4)--(1,4)--(1,5)--(0,5)--cycle;
\draw [line width=1.0, fill=red, fill opacity=.3] (4,0)--(5,0)--(5,3)--(2,3)--(2,2)--(4,2)--cycle;
\draw [line width=1.0, fill=green, fill opacity=.3] (0,1)--(3,1)--(3,2)--(2,2)--(2,3)--(0,3)--cycle;
\draw [line width=1.0, fill=blue, fill opacity=.3] (4,3)--(5,3)--(5,5)--(1,5)--(1,4)--(4,4)--cycle;
\end{scope}
\end{tikzpicture}
    \caption{Two districting plans of $5$ districts with $5$ blocks each applied to the voter distribution from Figure \ref{fig:square-grid}. }
    \label{fig:districting-plan-grid}
\end{figure}
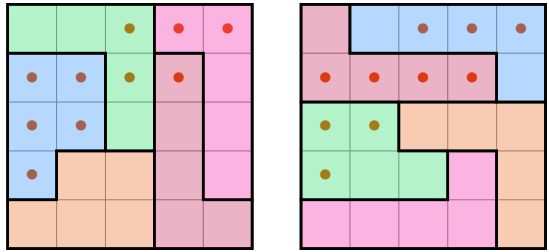

Figure \ref{fig:districting-plan-grid} details two possible districting plans for a $5\times 5$ grid. It also demonstrates that different districting plans can yield varying amounts of representation. The {\color{red}$\bullet$} party secures one seat with the plan on the left, as opposed to 3 seats with the plan on the right. \emph{Proportional representation} would call for $2$ seats won by the {\color{red}$\bullet$} party as there are $10$ {\color{red}$\bullet$} blocks. 

\newpage
\subsection{Methodology}
Harris provides an efficient algorithm for determining the various population distributions on an $n\times n$ square grid given sufficiently small dimensions \cite{Harris2010CountingIlk}. We use this algorithm to generate an exhaustive list of all $4006$ possible districting plans for a $5\times 5$ square grid, $\mathbb{D}$. There are a total of $2^{25}$ possible voter distributions $\Delta$ on a $5\times 5$ grid, with $2$ choices for each of the $25$ blocks. Then, taking the quotient by the dihedral group $D_4$ to account for symmetries of a square, we obtain $2^{23} \approx 8 \times 10^{6}$ possible districting plans. We were able to compute this exhaustive search over every districting plan in several hours on a standard laptop. The $6\times 6$ square grid case increases the computational complexity by a factor of over $200,000 < \frac{451206}{4006} \cdot \frac{2^{36}}{2^{35}},$ which exceeds the computational capacity available to us and most institutions. 

For each voter distribution that we iterate through, we record the following pieces of information: the number of {\color{red}$\bullet$} blocks, the clustering score ($\mathsf{Clus}$ and $\mathsf{ClusP}$), the distribution of the number of seats this voter distribution yields over all districting plans, and the expected (average) seats this voter distribution yields over all districting plans. 

\begin{Def}
We assume there are two parties, Dot ({\color{red}$\bullet$}) and Blank. Define $\mathsf{Num}$ as the number of dot-voting blocks in a given voting distribution $\Delta$.
\end{Def}

\subsection{Observations}

We use our exhaustive data from the $5\times5$ grid search to set a certain number of {\color{red}$\bullet$} blocks as a constant and retrieve all such voter distributions $\Delta$ with that constant number of {\color{red}$\bullet$} blocks. Using these, we examine the correlation between clustering $\mathsf{ClusP}(\Delta)$ and expected representation $\mathbb{E}(\mathsf{Rep}_\Delta)$ for that specific $\mathsf{Num}$. Additionally, we quantify this correlation, which we find to be approximately linear, using a linear regression test. 

\begin{figure}[h]
        \centering
        \includegraphics[width=4in]{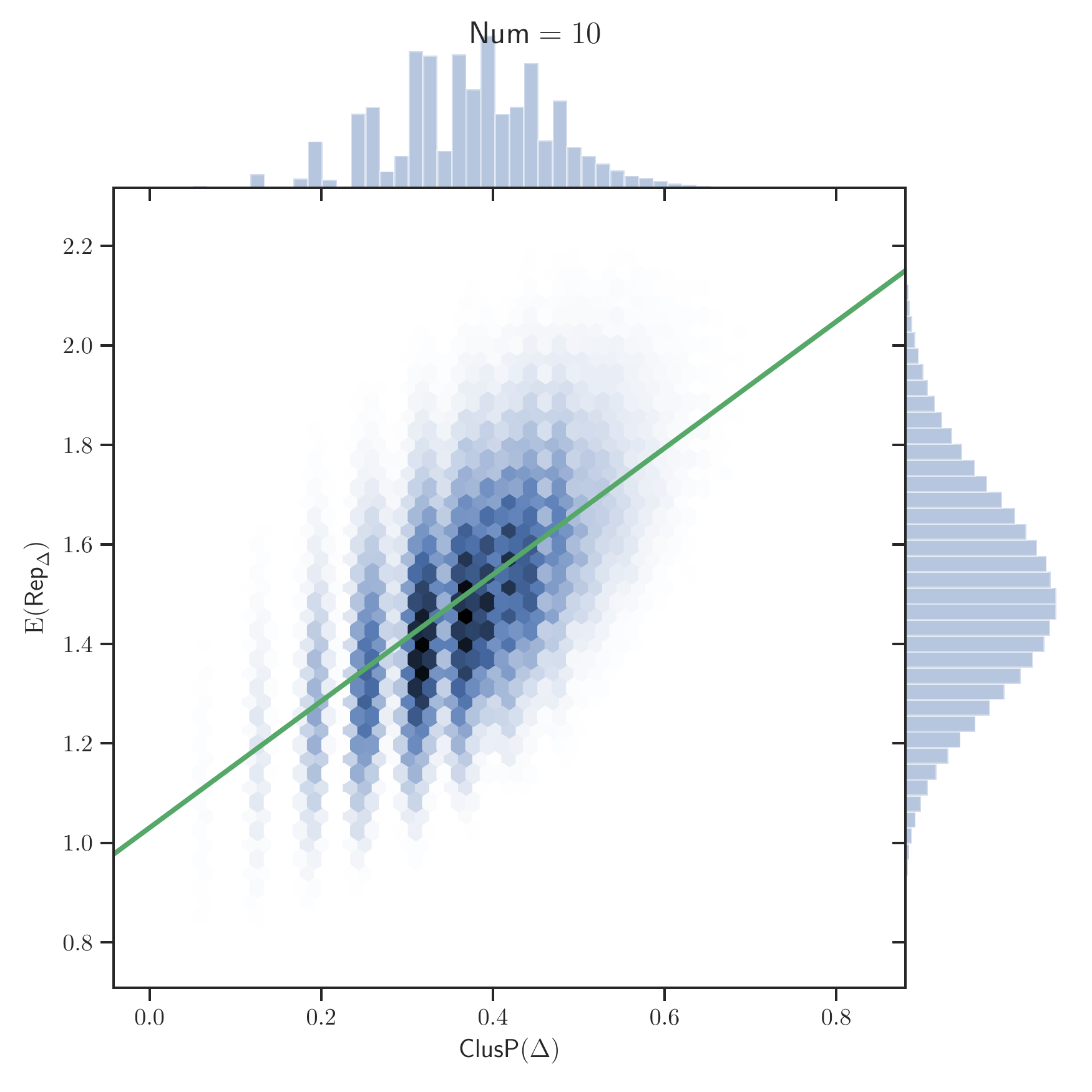}
        \caption{$\mathsf{ClusP}$ vs. $\mathbb{E}(\mathsf{Rep})$ for $\mathsf{Num} = 10$.}
        \label{fig:clusp10}
\end{figure}

For instance, Figure \ref{fig:clusp10} shows plot of $\mathsf{ClusP}(\Delta)$ against $\mathbb{E}(\mathsf{Rep}_\Delta)$ with a linear regression line of best fit superimposed. There are $\binom{25}{\mathsf{Num}}$ possible ways to arrange $\mathsf{Num}$ on a $5\times5$ grid, so a scatter plot obfuscates the relative densities of data points. Hence, we use a a heatmap-bin plot to display the data, with histograms of each data set, $\mathsf{ClusP}(\Delta)$  and $\mathbb{E}(\mathsf{Rep}_\Delta)$. The darker shades of the plot indicate higher relative densities. 

Furthermore, we can vary $\mathsf{Num}$ to extend these results, which each gives us a scatterplot for each $\mathsf{Num}$. We can find the slope of the linear regression line for each of these cases, and we can compare this slope with each corresponding $\mathsf{Num}$ value. Note that when $\mathsf{Num} = 0$, there is no possible $\Delta$ to calculate $\mathsf{ClusP}(\Delta)$ from. When $\mathsf{Num} = 1\text{ or }2$, the state space of possible voter distributions is so small that the linear regression test results in a slope of $0$.  See Table \ref{tab:LinRegTestNumH312} for the slope of every linear regression line for when $\mathsf{Num} < 13.$ Plots for other $\mathsf{Num}$ can be found in Appendix \ref{appendix:clush-v-rep}, which produces the results seen in Table \ref{tab:LinRegTestNumH312}.

\begin{table}[h]
\begin{tabular}{@{}rl@{}}
\toprule
$\mathsf{Num}$ & \textbf{Slope} \\ \midrule
1    & 0 \\
2    & 0 \\
3    & 0.2993106942 \\
4    & 0.6704477756 \\
5    & 1.040404652  \\
6    & 1.350056768  \\
7    & 1.553744171  \\
8    & 1.619139053  \\
9    & 1.527148112  \\
10   & 1.271799538  \\
11   & 0.8600922591 \\
12   & 0.3117844548 \\ \bottomrule \\
\end{tabular}
\caption{Results of linear regression test for $\mathsf{Num} = 3$ to $12$}
    \label{tab:LinRegTestNumH312}
\end{table}

\begin{figure}[h]
    \centering
    \begin{tikzpicture}

\definecolor{color0}{rgb}{0.12156862745098,0.466666666666667,0.705882352941177}

\begin{axis}[
tick align=outside,
tick pos=left,
x grid style={white!69.01960784313725!black},
xlabel={$\mathsf{Num}$},
xmajorgrids,
xmin=-0.207346698850806, xmax=26.2073466988508,
xtick style={color=black},
y grid style={white!69.01960784313725!black},
ylabel={$\mathbb{E}(\mathsf{Rep}_\Delta)/\mathsf{ClusP(\Delta)}$},
ymajorgrids,
ymin=-3.52474189883485, ymax=1.87485906983485,
ytick style={color=black}
]
\addplot [only marks, draw=color0, fill=color0, colormap/viridis]
table{%
x                      y
1 0
2 0
3 0.2993106942
4 0.6704477756
5 1.040404652
6 1.350056768
7 1.553744171
8 1.619139053
9 1.527148112
10 1.271799538
11 0.8600922591
12 0.3117844548
13 -0.3409122367
14 -1.05378706
15 -1.771325278
16 -2.427807723
17 -2.949118928
18 -3.255860772
19 -3.269021882
20 -2.921038193
21 -2.179352121
22 -1.102771422
23 0
24 0
25 0
};
\addlegendentry{Slope}
\end{axis}

\end{tikzpicture}
    \caption{This is a graph of the extended table of Table \ref{tab:LinRegTestNumH312}, which plots $\mathsf{Num}$ against the $\mathsf{ClusP}$ and $\mathbb{E}(\mathsf{Rep}_\Delta)$ score. This gives quantitative evidence that clustering is beneficial for minorities and detrimental to majorities. }
    \label{fig:numh-vs-slope}
\end{figure}
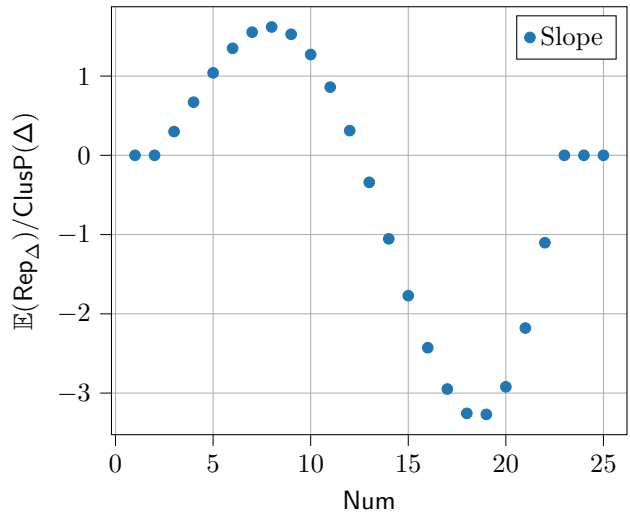

Then, we take the slope from each linear regression test and plot $\mathsf{Num}$ against these slope values. We display the results of this scatter plot in Figure \ref{fig:numh-vs-slope}. Figure \ref{fig:numh-vs-slope} shows that when Dots are in the minority such that $3 \leq \mathsf{Num} \leq 12,$ the slope is positive. Thus, there is a positive correlation between increased clustering, according to a higher clustering score, and increased representation. Therefore, if the minority desires maximal representation, then clustering is a better strategy. This correlation peaks at $\mathsf{Num}=8$ before decreasing in correlation as $\mathsf{Num}$ increases. Furthermore, when Dots are in the majority such that $13 \leq \mathsf{Num} \leq 22$, the slope is negative. This means that when the majority clusters itself, the expected representation reduces, so the majority's best strategy is to stop clustering and disperse evenly instead. For $\mathsf{Num} = 23,24$, the cases are symmetric to $\mathsf{Num} = 1,2$, with a very limited number of possible $\Delta$ and a linear regression test of slope $0$. 

Using our data, we can also determine the \textbf{absolute} best and worst voter distributions that yield the highest and lowest amounts of representation averaged over all the districting plans. We can make observations regarding these best and worst performing voter distributions. 

\begin{figure}[h]
\label{fig:square-gridgood}
    \centering
    \begin{tikzpicture}[scale=.65]
    \begin{scope}
\foreach \x in {0,...,5} {\draw [gray!85] (\x,0)--(\x,5);}
\foreach \y in {0,...,5} {\draw [gray!85] (0,\y)--(5,\y);}
\draw [line width=1.0] (0,0)--(5,0)--(5,5)--(0,5)--cycle;
\foreach \p/\q in {2/1, 3/1, 3/2, 4/1, 4/2, 4/3, 5/3, 5/4, 5/5}  {\node at (\p-.5,\q-.5) {\color{red}$\bullet$};}

\end{scope}

\begin{scope}[xshift = 6cm]
\foreach \x in {0,...,5} {\draw [gray!85] (\x,0)--(\x,5);}
\foreach \y in {0,...,5} {\draw [gray!85] (0,\y)--(5,\y);}
\draw [line width=1.0] (0,0)--(5,0)--(5,5)--(0,5)--cycle;
\foreach \p/\q in {2/1, 3/1, 3/2, 4/1, 4/2, 4/3, 4/4, 5/3, 5/4, 5/5}  {\node at (\p-.5,\q-.5) {\color{red}$\bullet$};}

\end{scope}

\begin{scope}[xshift = 12cm]
\foreach \x in {0,...,5} {\draw [gray!85] (\x,0)--(\x,5);}
\foreach \y in {0,...,5} {\draw [gray!85] (0,\y)--(5,\y);}
\draw [line width=1.0] (0,0)--(5,0)--(5,5)--(0,5)--cycle;
\foreach \p/\q in {2/1, 3/1, 3/2, 3/5, 4/1, 4/3, 4/4, 4/5, 5/2, 5/3, 5/5}  {\node at (\p-.5,\q-.5) {\color{red}$\bullet$};}

\end{scope}
\end{tikzpicture}

\vspace{1em}

\def\arraystretch{1.25}
\begin{tabular}{r|c|c|c}
$\mathbf{\mathsf{Num}}$                                   & $\mathbf{9}$ & $\mathbf{10}$ & $\mathbf{11}$ \\
\hline
$\mathbf{\mathsf{ClusP}(\Delta)}$                 & $0.621$      & $0.667$       & $0.514$       \\
\hline
$\mathbf{\mathbb{E}(\mathsf{Rep}_\Delta)}$                 & $2.015$      & $2.316$       & $2.601$       \\
\hline
$\mathbf{\mathsf{Var}(\mathsf{Rep}_\Delta)}$               & $0.247$      & $0.239$       & $0.260$       \\
\end{tabular}

    \caption{$5\times 5$ grids of optimal voter distribution with $\mathsf{Num} = 9,10,11$, left to right.}
    \label{fig:square-gridgood}
\end{figure}
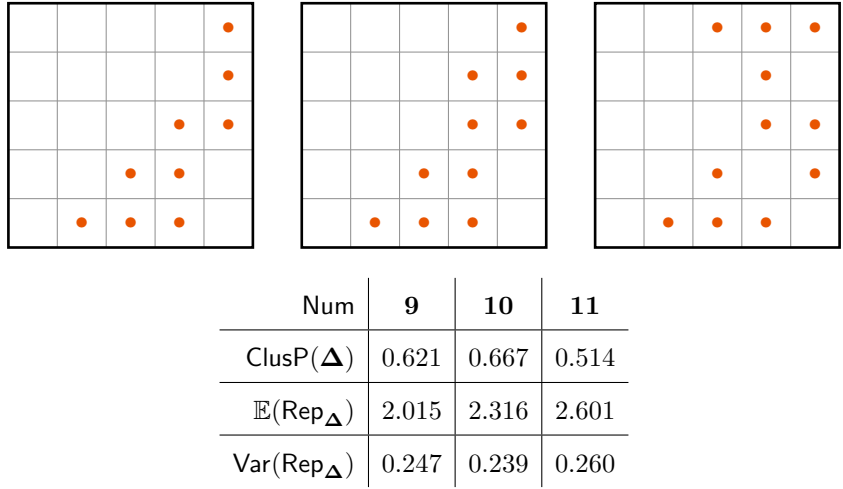

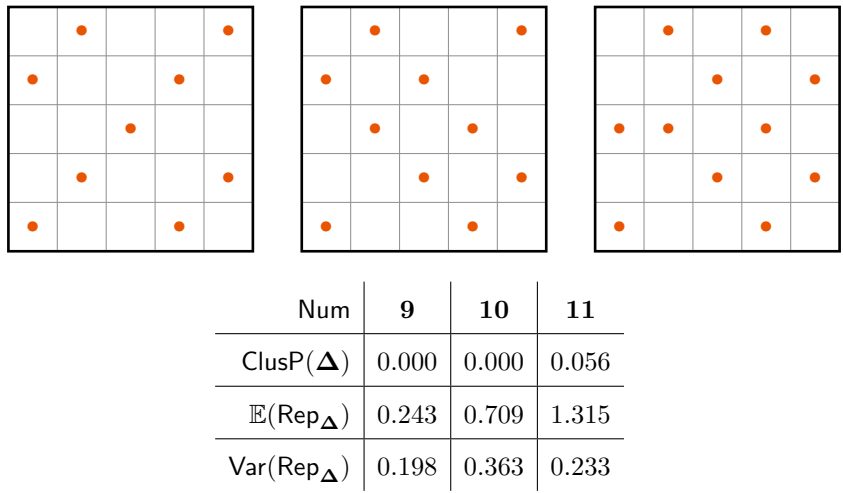
\begin{figure}[h]
\label{fig:square-gridbad}
    \centering
    \begin{tikzpicture}[scale=.65]
    \begin{scope}
\foreach \x in {0,...,5} {\draw [gray!85] (\x,0)--(\x,5);}
\foreach \y in {0,...,5} {\draw [gray!85] (0,\y)--(5,\y);}
\draw [line width=1.0] (0,0)--(5,0)--(5,5)--(0,5)--cycle;
\foreach \p/\q in {1/1, 1/4, 2/2, 2/5, 3/3, 4/1, 4/4, 5/2, 5/5}  {\node at (\p-.5,\q-.5) {\color{red}$\bullet$};}

\end{scope}

\begin{scope}[xshift = 6cm]
\foreach \x in {0,...,5} {\draw [gray!85] (\x,0)--(\x,5);}
\foreach \y in {0,...,5} {\draw [gray!85] (0,\y)--(5,\y);}
\draw [line width=1.0] (0,0)--(5,0)--(5,5)--(0,5)--cycle;
\foreach \p/\q in {1/1, 1/4, 2/3, 2/5, 3/2, 3/4, 4/1, 4/3, 5/2, 5/5}  {\node at (\p-.5,\q-.5) {\color{red}$\bullet$};}

\end{scope}

\begin{scope}[xshift = 12cm]
\foreach \x in {0,...,5} {\draw [gray!85] (\x,0)--(\x,5);}
\foreach \y in {0,...,5} {\draw [gray!85] (0,\y)--(5,\y);}
\draw [line width=1.0] (0,0)--(5,0)--(5,5)--(0,5)--cycle;
\foreach \p/\q in {1/1, 1/3, 2/3, 2/5, 3/2, 3/4, 4/1, 4/3, 4/5, 5/2, 5/4}  {\node at (\p-.5,\q-.5) {\color{red}$\bullet$};}

\end{scope}
\end{tikzpicture}

\vspace{1em}

\def\arraystretch{1.25}
\begin{tabular}{r|c|c|c}
$\mathbf{\mathsf{Num}}$                                   & $\mathbf{9}$ & $\mathbf{10}$ & $\mathbf{11}$ \\
\hline
$\mathbf{\mathsf{ClusP}(\Delta)}$                 & $0.000$      & $0.000$       & $0.056$       \\
\hline
$\mathbf{\mathbb{E}(\mathsf{Rep}_\Delta)}$ & $0.243$      & $0.709$       & $1.315$       \\
\hline
$\mathbf{\mathsf{Var}(\mathsf{Rep}_\Delta)}$               & $0.198$      & $0.363$       & $0.233$       \\
\end{tabular}
    \caption{$5\times 5$ grids of worst voter distribution with $\mathsf{Num} = 9,10,11$, left to right.}
    \label{fig:square-gridbad}
\end{figure}

Clustering leads to packing for the minority party, which tends to yield higher expected representation across all possible districting plans. We observe trends across all outlier cases that maximize the expected representation. In Figure \ref{fig:square-gridgood}, we show the representation-maximizing voter distributions for Dot vote share of 9, 10, and 11 out of 25. Note that although a higher clustering score generally correlates with higher expected representation, the districting plan that maximizes expected representation does not actually have the highest clustering score. An important observation is that these optimal voter distributions all contain \emph{enclaves} of blocks of the opposing party. These enclaves mean that the opposing party is the local minority and is far away from the majority of their blocks, which is sub-optimal for the opposing party, as it is difficult for a district contain enough blocks of the opposing party to win representation. This means that the votes secured for the opposing party by these `enclaves' are almost always wasted. 

We also observe trends across all outliers that minimize expected representation. Figure \ref{fig:square-gridbad} displays the representation-minimizing voter distributions for Dot vote share of 9, 10, and 11 out of 25. The least successful minority voter distributions have extremely low cluster scores, with each minority block adjacent to mostly or only majority blocks. The lack of minority clustering allows the opposing (majority) party to easily crack minority votes and secure maximal representation. 

This leads us to make the following conjecture: 

\begin{Conj}
\label{conj:clustering-relation}
	\textbf{(Clustering Relation).} For minority $\mathsf{Num}$, there is a positive correlation between $\mathsf{ClusP}$ and $\mathbb{E}(\mathsf{Rep}_\Delta)$. 
\end{Conj}

\begin{figure}[h]
    \centering
    \includegraphics[width = 6in]{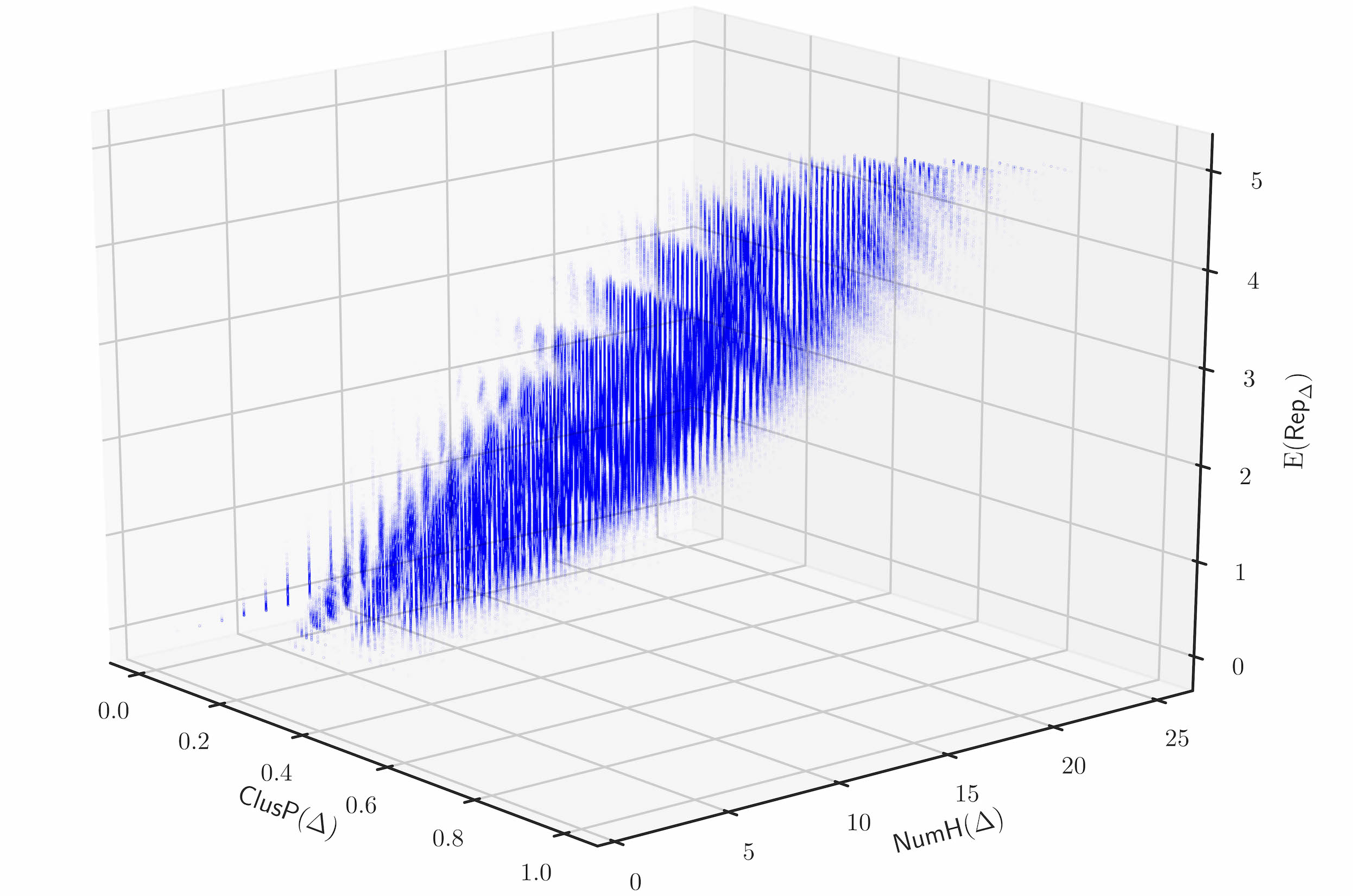}
    \caption{$\mathsf{Num}$ vs. $\mathsf{ClusP}(\Delta)$ vs. $\mathbb{E}(\mathsf{Rep}_\Delta).$}
    \label{fig:3dscatter}
\end{figure}

Figure \ref{fig:3dscatter} plots voter distribution, clustering score for Dots, and expected representation together on a three-dimensional plot. Figure \ref{fig:3dscatter} shows how expected representation varies with both clustering and vote share. It is clear from the plot that both increased clustering and increased vote share increase expected representation.

\newpage
\section{Metaheuristics for Optimization}
\label{sec:metaheuristics}

Recall that our goal is to find or estimate the best voter distribution that maximizes the representation of a certain political party within a geographic region split up into districts. The problem is two-fold: first, there must be a way to assign a score to a voter distribution; second, there must be an algorithm to efficiently find such a maximum within the large and discrete set of all voter distributions. 

\subsection{Evaluative function} 
\label{subsection:eval}
We tackle the first problem using a known algorithm. Our score for each voter distribution is $\mathbb{E}(\mathsf{Rep}_\Delta)$, which gives the expected representation over all possible districting plans for a voter distribution. A higher score represents a voter distribution that is robust against gerrymandering, in the sense that it is difficult and improbable to create a districting plan that grossly disadvantages the party in question. This measure also aligns with the `mean test' that the supreme court proposes. However, it is hard to compute this measure through an enumeration of all districting plans. Even for an $n \times n$ grid of squares, exhaustive enumeration of, or even counting, all possible partitions into $n$ equal-sized districts is computationally infeasible above $n=9$ as demonstrated in Table \ref{tab:polyomino-tilings}. Duchin gives and demonstrates a Metropolis-Hastings algorithm (also known as Markov Chain Monte Carlo) to stratify and uniformly sample over all districting plans \cite{Duchin2018GerrymanderingBaseline}. We denote this stochastic algorithm as the evaluative function $\mathsf{Eval}(\Delta)$ which approximates $\mathbb{E}(\mathsf{Rep}_\Delta)$. However optimized this algorithm is, note that it is still the time-limiting step in our proposed algorithms. 

Hence, we turn to the second problem, to find a voter distribution yielding maximal representation (or one which is good enough) within a large and discrete space of all voter distributions. A crude method to achieve this would simply be to generate random voter distribution, record down the respective representation that each yields, and have the algorithm return the voter distribution that yielded the highest representation after a set amount of tries. This is the \emph{Random algorithm}, and we use it as a benchmark to evaluate our proposed algorithms. One strategy we use to tackle this problem is through a quasi-greedy algorithm that mutates a voter distribution at every step instead of generating a completely new and random one. 

\subsection{Cellular automata}
\emph{Cellular automata} are discrete models in which the state of each cell affects its neighboring cells. We apply a cellular automata algorithm to our $5 \times 5$ grid to simulate clustering.

First, we define a condition for each block that determines if it is \emph{happy} or \emph{unhappy}. 

\begin{Def}
A block $b\in V$ is \emph{happy} when the proportion of its like neighbors over its total number of neighbors (i.e. those it shares an edge with) is above a set threshold $\theta$. Likewise, the block $b$ is \emph{unhappy} when the proportion of its like neighbors over its total number of neighbors is less than $\theta$. 
\end{Def}

The algorithm proposes an evolution of our voter distribution $\Delta$ by taking the set of all the unhappy blocks (which could be both empty and {\color{red}$\bullet$} party blocks) and shuffling the blocks (in essence, permuting the individual blocks). Notably, this preserves $\mathsf{Num}$ and achieves an effect of increasing the clustering score. This is due to the fact that the blocks we are perturbing contribute to a lower clustering score (as they have a lower number of like neighbors), so we are decreasing the number of `unlike' connections. Its implementation is given in Algorithm \ref{algorithm:cellular-automata}.

\begin{algorithm}[H]
    \caption{Cellular automata evolution}
    \label{algorithm:cellular-automata}
    \begin{algorithmic}[1] 
    \Require
      \Statex Dual graph $G = (V, E)$, discrete voter distribution function $\Delta: V\to \{0, 1\}$, and threshold $\theta\in [0,1]$.
        \Procedure{Evolve}{$\Delta$} \Comment{Initial voter distribution $\Delta$}
            \State $S \gets \{\}$ \Comment{Set of blocks $b$ with proportion $<\theta$ of similar neighbors}
            \State $\{\Delta\}_{\mathrm{swap}} \gets \{\}$ \Comment{$\{\Delta\}_{\mathrm{swap}} = \{\Delta(b)\ |\ b\in S\}$}
            \For{$b$ in $V$}
            	\State $conn_{total} \gets 0$
            	\State $conn_{\mathrm{same}} \gets 0$
            	\For{$b_n$ in $\{b_n\ |\ (b, b_n)\in E\}$} \Comment{For all blocks adjacent to $b$}
            	\IfThen{$\Delta(b) = \Delta(b_n)$}{$conn_{\mathrm{same}} \gets conn_{\mathrm{same}} + 1$} \Comment{\# of similar connections}
            	\State $conn_{\mathrm{total}} \gets conn_{\mathrm{total}} + 1$ \Comment{\# of total connections}
            	\EndFor \Comment{After enumeration of $b_n$}
            	            	\If{$conn_{\mathrm{same}}/conn_{\mathrm{total}} < \theta$}
            	\State $S \gets S \cup \{b\}$
            	\State \textbf{append} $\Delta(b) \to \{\Delta\}_{\mathrm{swap}}$, preserving order
            	\EndIf
            \EndFor \Comment{After enumeration of $b$ in $V$}
            \State \textbf{shuffle} $\{\Delta\}_{\mathrm{swap}}$ \Comment{Permutes indices of $\Delta_i\in\{\Delta\}_{\mathrm{swap}}$}
            \State \textbf{define} $\Delta'(b) = \begin{cases}
            \Delta_i\text{ in }\{\Delta\}_{\mathrm{swap}} & \textbf{for } b = b_i\in S \\
            \Delta(b) & \textbf{otherwise } (\text{i.e. }b\not\in S)
            \end{cases}$
            \State \Return $\Delta'$ \Comment{Returns mutated voter distribution function $\Delta'$}
        \EndProcedure
    \end{algorithmic}
\end{algorithm}

The progression below gives a visual example of what the cellular automata algorithm does in each iteration. 

\begin{figure}[h]
\label{fig:cellular-automata}
    \centering
    \begin{tikzpicture}[scale=.43]
    \begin{scope}
        \foreach \x in {0,...,9} {\draw [gray!85] (\x,0)--(\x,9);}
        \foreach \y in {0,...,9} {\draw [gray!85] (0,\y)--(9,\y);}
        \foreach \p/\q in {9/2, 9/9, 8/6, 8/7, 7/1, 7/3, 7/4, 7/5, 7/6, 6/2, 6/3, 6/5, 6/6, 6/9, 4/1, 4/5, 4/6, 4/7, 4/8, 3/8, 3/9, 2/1, 2/2, 2/3, 2/4, 2/5, 2/7, 2/9, 1/1, 1/7}  {\node at (\p-.5,\q-.5) {\color{red}$\bullet$};}
        \draw [line width=1.0] (0,0)--(9,0)--(9,9)--(0,9)--cycle;
    \end{scope}
    \begin{scope}[xshift = 10cm]
        \foreach \x in {0,...,9} {\draw [gray!85] (\x,0)--(\x,9);}
        \foreach \y in {0,...,9} {\draw [gray!85] (0,\y)--(9,\y);}
        \foreach \r/\c in {9/2, 9/9, 8/7, 7/1, 7/2, 6/1, 6/2, 6/4, 6/9, 4/1, 4/5, 4/9, 3/1, 3/7, 2/5, 2/7, 2/8, 2/9, 1/2, 1/7}  {\draw [fill=orange!20] (\r,\c)--(\r -1,\c)--(\r -1,\c -1)--(\r,\c -1)--cycle;}
        \foreach \p/\q in {9/2, 9/9, 8/6, 8/7, 7/1, 7/3, 7/4, 7/5, 7/6, 6/2, 6/3, 6/5, 6/6, 6/9, 4/1, 4/5, 4/6, 4/7, 4/8, 3/8, 3/9, 2/1, 2/2, 2/3, 2/4, 2/5, 2/7, 2/9, 1/1, 1/7}  {\node at (\p-.5,\q-.5) {\color{red}$\bullet$};}
        \draw [line width=1.0] (0,0)--(9,0)--(9,9)--(0,9)--cycle;
    \end{scope}
    \begin{scope}[xshift = 20cm]
        \foreach \x in {0,...,9} {\draw [gray!85] (\x,0)--(\x,9);}
        \foreach \y in {0,...,9} {\draw [gray!85] (0,\y)--(9,\y);}
        \foreach \r/\c in {9/2, 9/9, 8/7, 7/1, 7/2, 6/1, 6/2, 6/4, 6/9, 4/1, 4/5, 4/9, 3/1, 3/7, 2/5, 2/7, 2/8, 2/9, 1/2, 1/7}  {\draw [fill=orange!20] (\r,\c)--(\r -1,\c)--(\r -1,\c -1)--(\r,\c -1)--cycle;}
        \foreach \p/\q in {9/2, 9/9, 8/6, 7/1, 7/3, 7/4, 7/5, 7/6, 6/1, 6/2, 6/3, 6/4, 6/5, 6/6, 6/9, 4/1, 4/6, 4/7, 4/8, 3/1, 3/8, 3/9, 2/1, 2/2, 2/3, 2/4, 2/8, 2/9, 1/1, 1/7}  {\node at (\p-.5,\q-.5) {\color{red}$\bullet$};}
        \draw [line width=1.0] (0,0)--(9,0)--(9,9)--(0,9)--cycle;
    \end{scope}
    \begin{scope}[xshift=30cm]
        \foreach \x in {0,...,9} {\draw [gray!85] (\x,0)--(\x,9);}
        \foreach \y in {0,...,9} {\draw [gray!85] (0,\y)--(9,\y);}
        \foreach \r/\c in {9/2, 9/9, 8/6, 7/1, 7/2, 6/9, 5/1, 4/1, 4/6, 4/9, 2/4, 1/2, 1/7, 1/8}  {\draw [fill=orange!20] (\r,\c)--(\r -1,\c)--(\r -1,\c -1)--(\r,\c -1)--cycle;}
        \foreach \p/\q in {9/2, 9/9, 8/6, 7/1, 7/3, 7/4, 7/5, 7/6, 6/1, 6/2, 6/3, 6/4, 6/5, 6/6, 6/9, 4/1, 4/6, 4/7, 4/8, 3/1, 3/8, 3/9, 2/1, 2/2, 2/3, 2/4, 2/8, 2/9, 1/1, 1/7}  {\node at (\p-.5,\q-.5) {\color{red}\small$\bullet$};}
        \draw [line width=1.0] (0,0)--(9,0)--(9,9)--(0,9)--cycle;
    \end{scope}
    \end{tikzpicture}
        \caption{A $9\times9$ square grid with $\textsf{Num} = 30$ demonstrating the respective steps of the cellular automata evolution, with $\theta= 0.4$.}
\end{figure}
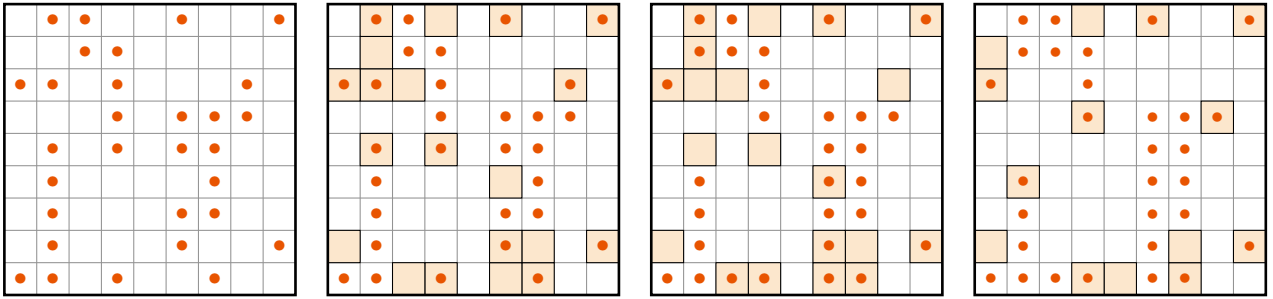

The first grid on the left of Figure \ref{fig:cellular-automata} is our initial voter distribution $\Delta$. We can identify all the unhappy tiles on our grid by enumerating the tiles where the proportion of like tiles connected to it is less than $0.4$. In our case, this means a tile with less than $2$ similar edges out of $4$ or $3$ total edges, or $0$ similar edges if it has $2$ total outgoing edges. Unhappy tiles are highlighted on the second grid. The algorithm then shuffles all the unhappy tiles around, leaving the happy tiles as is. This produces the diagram in the third grid, which still highlights the unhappy tile locations after being shuffled. Those blocks that are not highlighted have not been changed. The last grid demonstrates what re-evaluating the unhappy tiles in our new grid would look like. As you can see, the number of unhappy tiles has decreased. 

Relying on our conclusion in Conjecture \ref{conj:clustering-relation} and the fact that cellular automata generally gives a more clustered voter distribution with every evolution, we can say that the cellular automata algorithm also improves expected representation $\mathbb{E}(\mathsf{Rep}_\Delta)$ for the minority. Hence, we can use the cellular automata algorithm to approximate a quasi-greedy algorithm, which attempts to find similar voter distributions with better expected representation every time. This becomes useful for Algorithms \ref{algorithm:rrils} and \ref{algorithm:sa} detailed below, both of which rely on the fact that we can easily find or generate an optimal `neighbor' given some voter distribution $\Delta$. This is non-trivial, given that $\{\Delta\}$ is discrete and so large. 

\hfill
\begin{flushright}
\emph{(Continued on next page\dots)}
\end{flushright}
\newpage
\subsection{Random-restart iterated local search algorithm (RRILS)}
\label{subsection:rrils}
Using the cellular automata algorithm as a quasi-greedy algorithm allows us to apply this algorithm repeatedly to generate increasingly better voter distributions (hence the `iterated local search'). Of course, this algorithm is succeptible to terminating at a local maxima when we are trying to find the global maxima instead. We can randomly restart the algorithm at some given point any time we reach a local maxima, and this gives a relatively straightforward method for finding a stronger maxima. 

In words, the algorithm executes as follows: first, we set the number of trials as $k_\mathrm{max}$. The algorithm will then generate a random voter distribution, $\Delta$. It then approximates the $\mathbb{E}(\mathsf{Rep})$ of this $\Delta$ using the evaluative function $\mathsf{Eval}$ detailed in subsection \ref{subsection:eval}. Then using the greedy algorithm, which shuffles unhappy blocks by swapping them, RRILS can propose a greedy evolution using cellular automata. If there are no new evolutions to be proposed, another random initial voter distribution will be generated. This process repeats, and after iterating $k_\mathrm{max}$ times, the algorithm return the $\Delta$ with maximum $\mathbb{E}(\mathsf{Rep}_\Delta)$ that it has found.
\begin{algorithm}[H]
	\caption{Random-restart iterated local search}
	\label{algorithm:rrils}
	\begin{algorithmic}[1]
	\Require
	\Statex Evaluative function $\mathsf{Eval}: \{\Delta\} \to \mathbb{R}$ which approximates $\mathbb{E}(\mathsf{Rep}_\Delta)$. 
	\Procedure{RandomRestart}{$k_{\mathrm{max}}$}
	\State $k \gets 0$
		\While{$k < k_{\mathrm{max}}$}
			\State $\Delta \gets \mathrm{rand}(\Delta)$ \Comment{Generates a random initial voter distribution}
			\State $\Delta_{\mathrm{new}} \gets$ null \Comment{Always accepts initial $\Delta$}
			\While{$\Delta \neq \Delta_{\mathrm{new}}$}				
			\State $\mathrm{Trials}[\Delta]\gets \mathsf{Eval}(\Delta)$ \Comment{Appends $\mathsf{Eval}(\Delta)$ to dictionary \underline{Trials} with key $\Delta$}
				\State $\Delta_{\mathrm{new}} \gets \textsc{Evolve}(\Delta)$ \Comment{Proposes an evolution of $\Delta$}
				\State $k \gets k + 1$
			\EndWhile
		\EndWhile
		\State $\Delta_{\mathrm{max}} \gets \kappa$ such that $\mathrm{Trials}[\kappa]$ maximal
		\State \Return $\Delta_{\mathrm{max}}$ \Comment{Returns $\Delta$ that gives maximal $\mathsf{Eval}(\Delta)$}
	\EndProcedure
	\end{algorithmic}
\end{algorithm}

\subsection{Simulated annealing}
\label{subsection:sa}
\emph{Simulated Annealing} is another metaheuristic designed for optimization, meant for cases where the search space is discrete, such as the space of all voter distributions $\{\Delta\}$. 

The algorithm tries to achieve a balance between exploiting (descending a gradient to reach local extrema) such as an iterated local search algorithm, as well as exploring (sampling the search space completely randomly for optimal voter distributions). The algorithm begins with an initial temperature (parameter) of $T_0$. This is the maximum temperature it will ever be at and represents a tendency for the algorithm to explore rather than exploit, which means it readily accepts worse states of proposed voter distributions. With every iteration, the temperature $T$ becomes $T\cdot \alpha$ due to a cooling schedule $\alpha$, which lowers the temperature. As the temperature decreases, the likelihood of accepting a worse state also decreases (defined by a function $\mathsf{Prob}$). 

This first version of simulated annealing we present is modified to take into account a known greedy algorithm. Similar to the RRILS, we evaluate $\Delta$'s and propose an evolution of $\Delta$. The algorithm then accepts this proposed state with a probability $\mathsf{Prob}(\delta_{score},T)$ such that $\mathsf{Prob}(\delta_{score},T) = e^{\delta_\mathrm{score}/T}$. The algorithm then decreases the temperature (if it chooses to accept the proposed state), or accepts the random state with a probability of \emph{$\theta_r$} which is a predefined constant. This process repeats, and after iterating $k_\mathrm{max}$ times, the algorithm returns the $\Delta$ which maximizes $\mathbb{E}(\mathsf{Rep}_\Delta)$.

\begin{algorithm}[H]
	\caption{Simulated annealing}
	\label{algorithm:sa}	
	\begin{algorithmic}[1]
		\Require
	\Statex Evaluative function $\mathsf{Eval}: \{\Delta\} \to \mathbb{R}$ which approximates $\mathbb{E}(\mathsf{Rep}_\Delta)$, acceptance probability function $\mathsf{Prob}: \{\Delta\} \times \mathbb{R} \to [0, 1]\in\mathbb{R}$, and following hyper-parameters: initial temperature $T_0$, cooling schedule $\alpha \in [0, 1]$, threshold $\theta$ (of \textsc{Evolve}), and constant probability of accepting random state $\theta_r$. 
	\Procedure{SimulatedAnneal}{$k_{\mathrm{max}}$}
	\State $k \gets 0$
	\State $T \gets T_0$ \Comment{Begins at initial temperature $T_0$}
	\State $\Delta \gets \mathrm{rand}(\Delta)$
	\State $score \gets \mathsf{Eval}(\Delta)$
	\While{$k < k_{\mathrm{max}}$}
		\State $\mathrm{Trials}[\Delta]\gets score$ \Comment{Appends $\mathsf{Eval}(\Delta)$ to dictionary \underline{Trials} with key $\Delta$}
		\State $\Delta_{\mathrm{new}} \gets \textsc{Evolve}(\Delta)$ \Comment{Proposes an evolution of $\Delta$}
		\State $score_\mathrm{new} \gets \mathsf{Eval}(\Delta_\mathrm{new})$
		\State $\delta_\mathrm{score} \gets score_\mathrm{new} - score$
		\If{$\mathsf{Prob}(\delta_\mathrm{score}, T) > \mathrm{rand}(0, 1)$} \Comment{Accepts proposed state with $\mathsf{Prob}(\delta_\mathrm{score}, T)$}
			\State $\Delta \gets \Delta_\mathrm{new}$
			\State $score \gets score_\mathrm{new}$
			\State $T \gets T\cdot \alpha$ \Comment{Decreases temperature}
		\ElsIf{$\theta_r > \mathrm{rand}(0, 1)$}	 \Comment{Otherwise accepts random state with probability $\theta_r$}
			\State $\Delta \gets \mathrm{rand}(\Delta)$
			\State $score \gets \mathsf{Eval}(\Delta)$
		\EndIf
		\State $k \gets k + 1$
	\EndWhile
	\State $\Delta_{\mathrm{max}} \gets \kappa$ such that $\mathrm{Trials}[\kappa]$ maximal
	\State \Return $\Delta_{\mathrm{max}}$ \Comment{Returns $\Delta$ that gives maximal $\mathsf{Eval}(\Delta)$}
	\EndProcedure
	\end{algorithmic}
\end{algorithm}

\subsection{Random Simulated Annealing}
\label{subsection:rsa}
\emph{Random Simulated Annealing} works largely the same way as the aforementioned simulated annealing in Algorithm \ref{algorithm:sa}. However, the steps that this algorithm takes are no longer the greedy steps prescribed by the cellular automata evolution algorithm. Where Algorithm \ref{algorithm:sa} would call {\sc Evolve} on line 8, Random Simulated Annealing calls {\sc Step}. {\sc Step} takes a parameter $n$, the number of blocks to shuffle, randomly selects $n$ blocks, and shuffles their political preferences in place on the dual graph. This is very similar to what our cellular automata algorithm {\sc Evolve} does, but instead of selecting a set of blocks that fits a certain criteria to swap (in {\sc Evolve} we selected `unhappy' blocks), {\sc Step} just selects $n$ random blocks to swap. From our tests, we have found that it is best that the random threshold $\theta_r$ be set to $0$, which means that the algorithm will never select a totally random state. This is because the steps that the algorithm is making is already random and is not susceptible to being trapped at a local extrema point. 

One benefit of the random variant of simulated annealing is that it does not rely on the Clustering Conjecture, and hence, we can use this algorithm for both the minority and majority parties. 

\subsection{Benchmark Algorithm}
\label{subsection:benchmark}
We use a naive benchmark algorithm to test the performance of the proposed algorithms. The benchmark algorithm simply generates new voter distributions of a set $\mathsf{Num}$ and evaluates them at every step. It returns the voter distribution that has yielded the largest expected representation for a given $\mathsf{Num}$. 

\subsection{Results}
Using the 3 algorithms detailed in Subsections \ref{subsection:rrils}, \ref{subsection:sa}, and \ref{subsection:rsa}, we can evaluate the computational improvements that they provide over our benchmark algorithm (and the next best naive alternative), the random sampling in Subsection \ref{subsection:benchmark}. 

In these trials, we use the dual graph of the $5\times 5$ square grid as described in Section \ref{sec:grid}. Since we have exhaustively analyzed this case, we know the maximum outcomes for each $\mathsf{Num}$, which gives us yet another benchmark to compare the results to. We run simulations by stochastically sampling the performance of each algorithm a large number of times (here, we run 10,000 trials) and cap each $k_\mathrm{max}$ to a certain value. For each $k_\mathrm{max}$ value, we can determine the average outcome (the optimized maximum expected representation) of each algorithm. As $k_\mathrm{max}$ is the number of iterations our program makes (or the computation time of our program), the algorithm also produces results that are closer to the global maximum as it increases. 

\begin{landscape}
\begin{figure}[htp]
    \centering
        \vspace{-0.8em}
        \hspace{-2em}
    \noindent\makebox[\textwidth]{
    \includegraphics[scale=1.15, trim={1cm 0.5cm 2cm 1cm},clip]{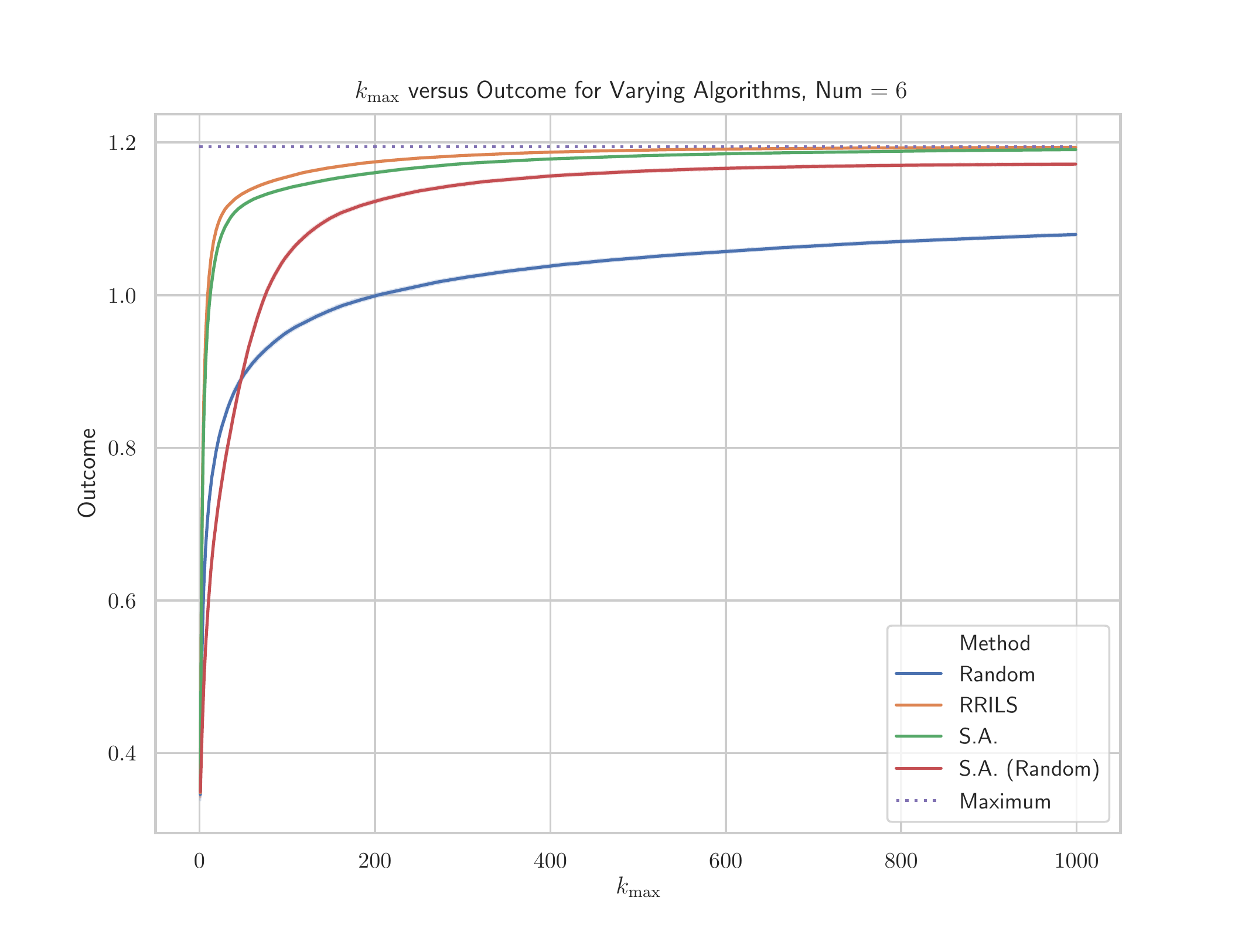}
    }
    \caption{The rates at which each algorithm approaches the absolute known maximum for a $5\times 5$ grid of $\mathsf{Num}=6$ for differing $k_\mathrm{max}$.}
    \label{fig:algorithm-comparison-6}
\end{figure}
\end{landscape}

Plotting out each of these results gives Figure \ref{fig:algorithm-comparison-6}, where the rates of each algorithm are visually displayed for $\mathsf{Num}=6$ and up to a $k_\mathrm{max}$ of 1000. For the case of $\mathsf{Num}=6$, the random-restart iterated local search algorithm performed similarly to simulated annealing, both of which performed better than the random variation of simulated annealing. Notably, all three algorithms asymptotically approached the absolute maximum significantly faster than random selection. 

\begin{landscape}
\begin{figure}[htp]
    \centering
        \vspace{-0.8em}
        \hspace{-2em}
    \noindent\makebox[\textwidth]{
    \includegraphics[scale=1.15, trim={1cm 0.5cm 2cm 1cm},clip]{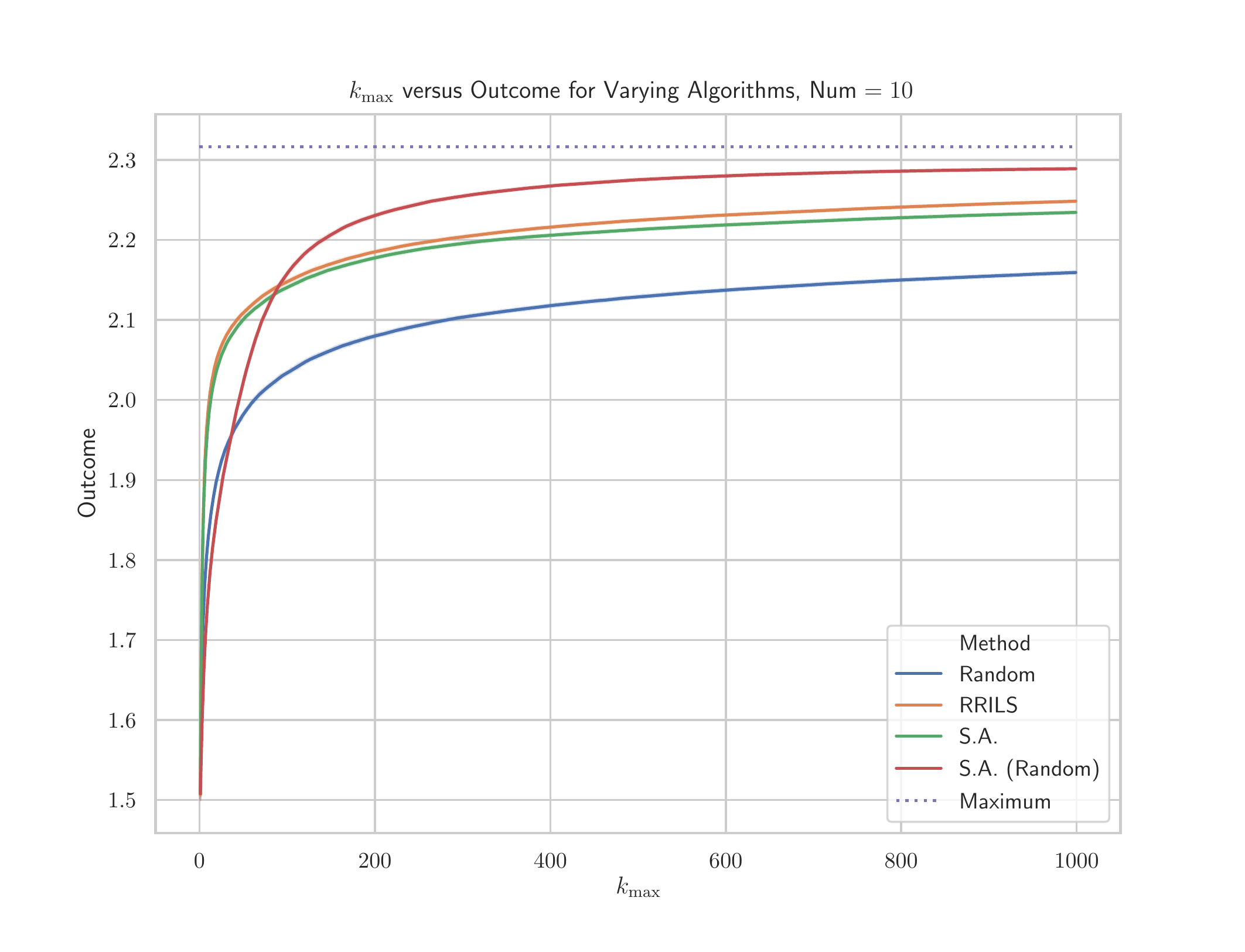}
    }
    \caption{The rates at which each algorithm approaches the absolute known maximum for a $5\times 5$ grid of $\mathsf{Num}=10$ for differing $k_\mathrm{max}$.}
    \label{fig:algorithm-comparison-10}
\end{figure}
\end{landscape}

However, when we change $\mathsf{Num}$ to equal $10$, comparing the algorithms gives us different results. Figure \ref{fig:algorithm-comparison-10} is a variation on Figure \ref{fig:algorithm-comparison-6}, where $\mathsf{Num}=6$. In this case, the random variation of simulated annealing performs better than both random-researt iterated local search and the original greedy variant of simulated annealing. 

Unfortunately, we can only conclude that our algorithms differ on a case-by-case basis and that it requires additional evaluation to determine the best algorithm for such an optimization. 

\section{Conclusion and Further Research}
By performing an exhaustive search all possible voter distributions and districting plans of a $5\times5$ grid, we were able to observe that a positive correlation exists between clustering and expected representation, giving us statistical evidence that clustering is beneficial for the minority. We also analyzed outlier cases of voter distribution that give maximum and minimum representation for a certain Dot vote share, discovering trends that indicate clustering generally leads to a larger amount of expected representation. This observation specifically extends and disproves, for certain cases, the observations of Chen, Rodden \cite{Chen2013UnintentionalLegislatures}. We use this result to implement the cellular automata, random-restart iterated local search, and simulated annealing algorithms to optimize the voter distribution for the highest expected representation. 

For large $n$, starting with $n > 5$, it becomes much harder to exhaustively search all districting plans and voter distributions. We hope to generalize many of our theorems and findings to larger grids, as well as other uniform geometries such as triangular and hexagonal tilings, as well as general dual graphs. 

An important direction of future gerrymandering research may be to determine methods which give representative random samples of a voter distribution given a particular districting plan or voter distribution. While MCMC already allows mathematicians to search the metagraph of districting plans using a random walk, there is no equivalent method to random walk on the space of all voter distributions in a way that provides a representative sample (the search space of all voter distributions is much larger than the search space of all districting plans). 

Extensive evidence also shows that the distribution of seats won over varying districting plans for a given population distribution is approximately normal, and we believe the central limit theorem could be leveraged to prove this observation. This allows us to use implications of normal distributions to our advantage when analyzing the effectiveness or result of a given voter distribution. 

We hope that our research serves as insight into the benefits of clustering for minority parties and as a basis for future research into exhaustive searches on other naive grids. We also hope that it gives insight into patterns and quantifications of specific voter distributions that do not rely on stratified MCMC searches to estimate the seats distribution. 







\section*{Acknowledgements}
The results in this paper originated from a research project at PROMYS 2019. We are deeply grateful to Diana Davis for proposing the problem, as well as for her constant support and guidance. We thank our counselor Kenz Kallal for his mentoring and for always being there to support us. This research would not be possible without David Fried, Glenn Stevens, Roger Van Peski, the PROMYS Foundation, and the Clay Mathematics Institute. We would also like to thank Moon Duchin for her ideas, suggestions, and insight into this topic. Additionally, the first author would like to thank Justin Almeida for his guidance and support following the end of the program. 


\printbibliography

\newpage
\appendix
\section{Computation}
The code used to perform our search, as well as the code for our metaheuristic algorithms (and their evaluations), can be found at \url{https://github.com/jiahuac/GerryGrid}. 

\section{$\mathsf{ClusP}$ Against $\mathbb{E}(\mathsf{Rep})$ Given $\mathsf{Num}$}
\label{appendix:clush-v-rep}
\centering
We include the $\mathsf{ClusP}$ vs. $\mathbb{E}(\mathsf{Rep})$ graphs for varying $\mathsf{Num}$ for a $5\times 5$ grid. 

        \includegraphics[width=5cm]{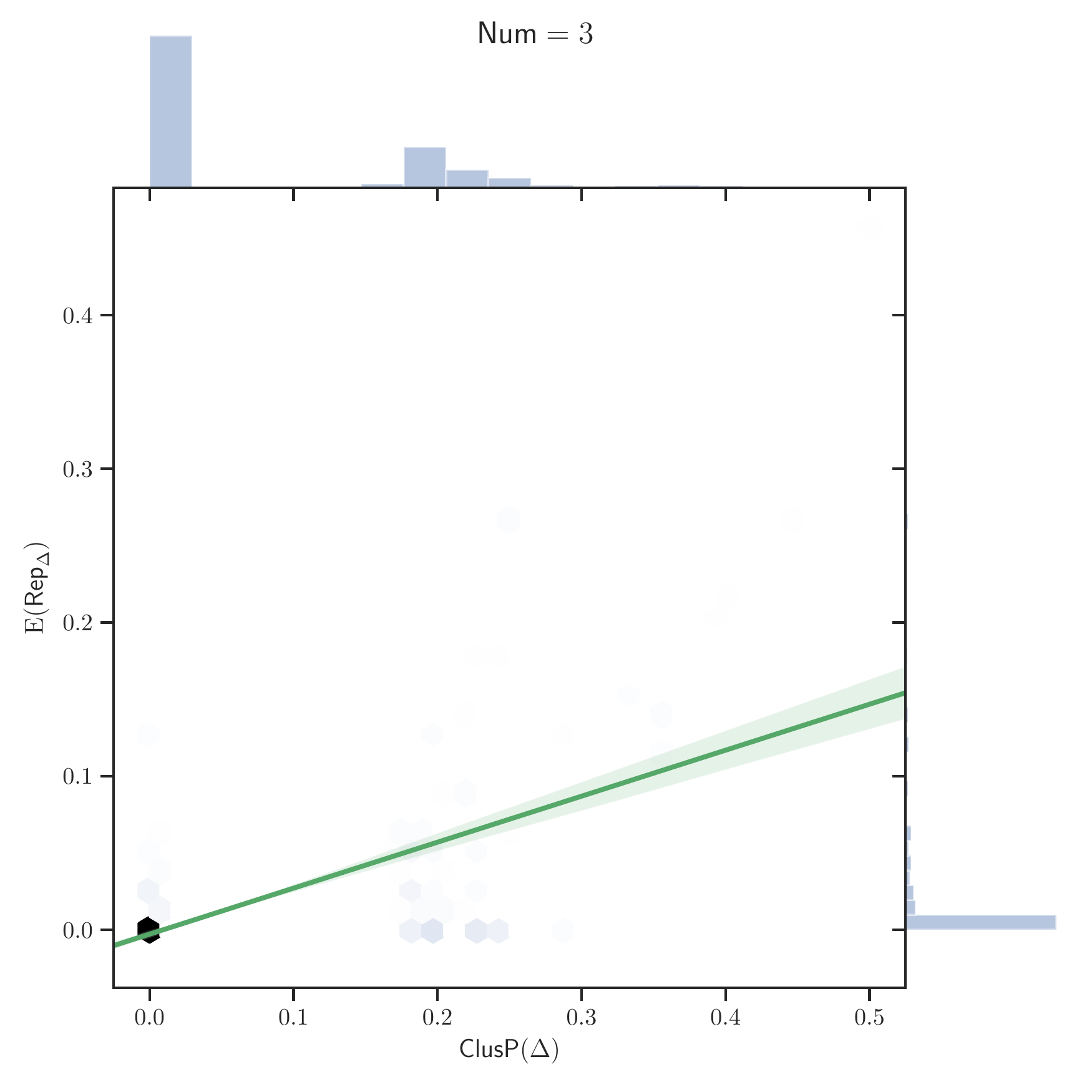}
        \includegraphics[width=5cm]{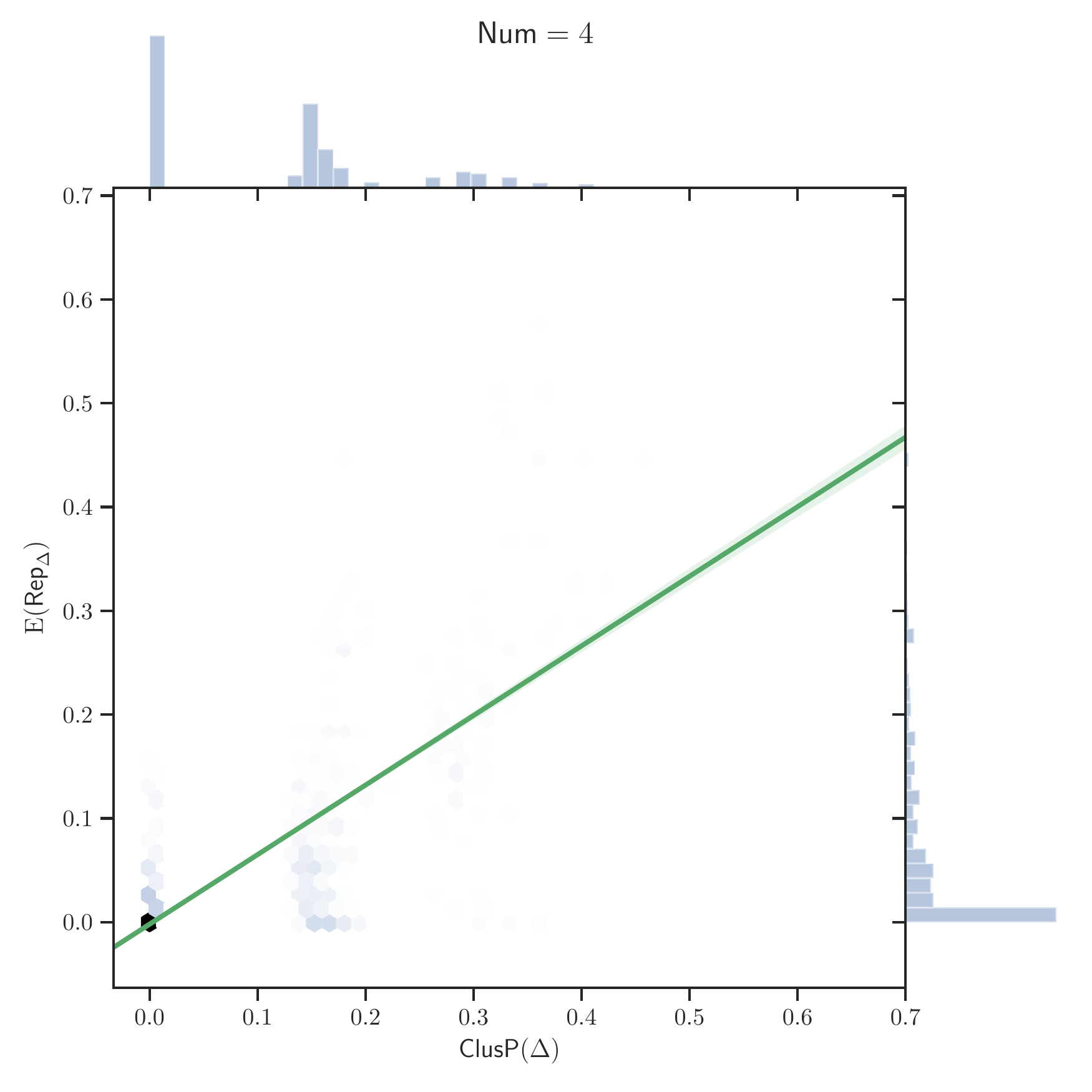}
        \includegraphics[width=5cm]{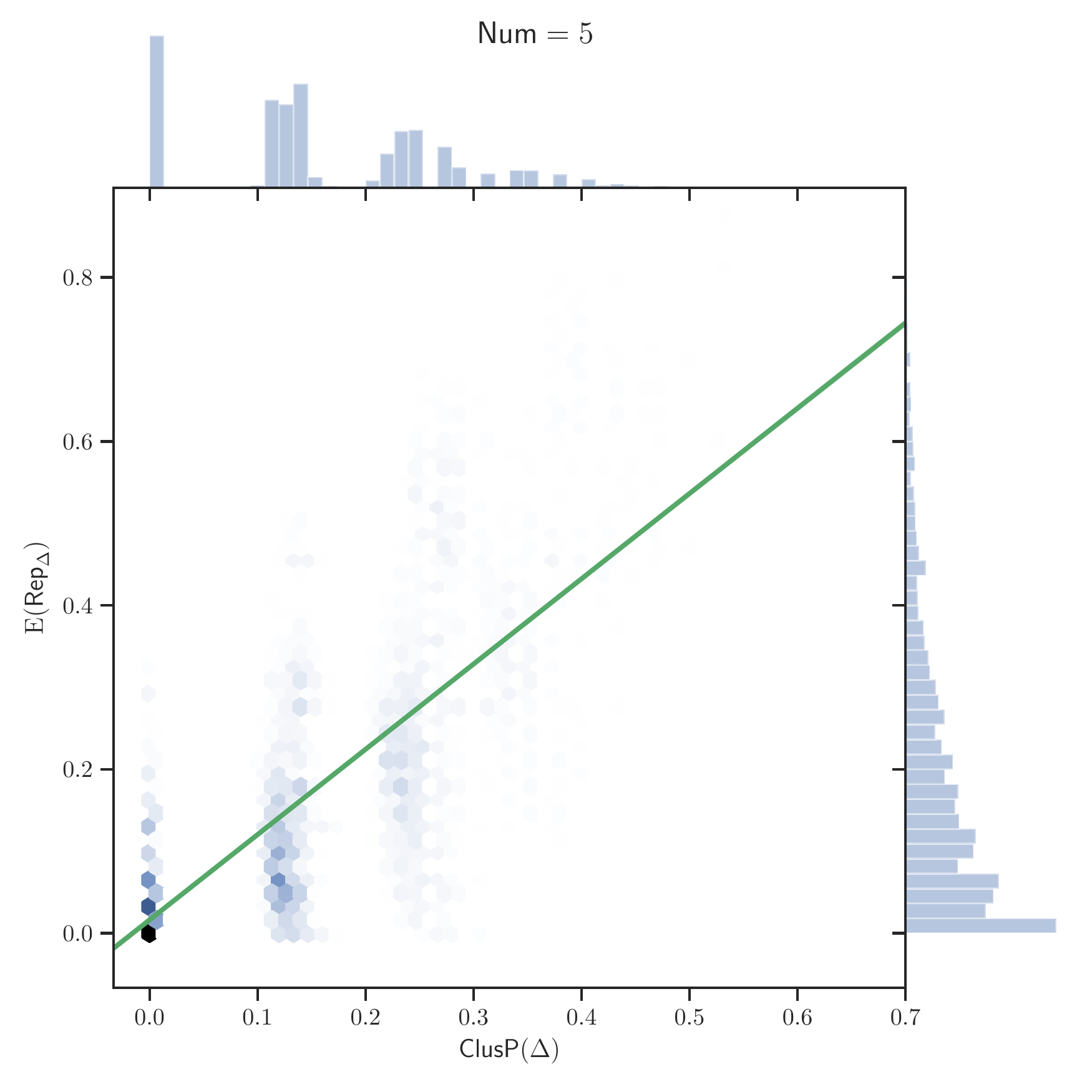}
        \\[1em]
        \includegraphics[width=5cm]{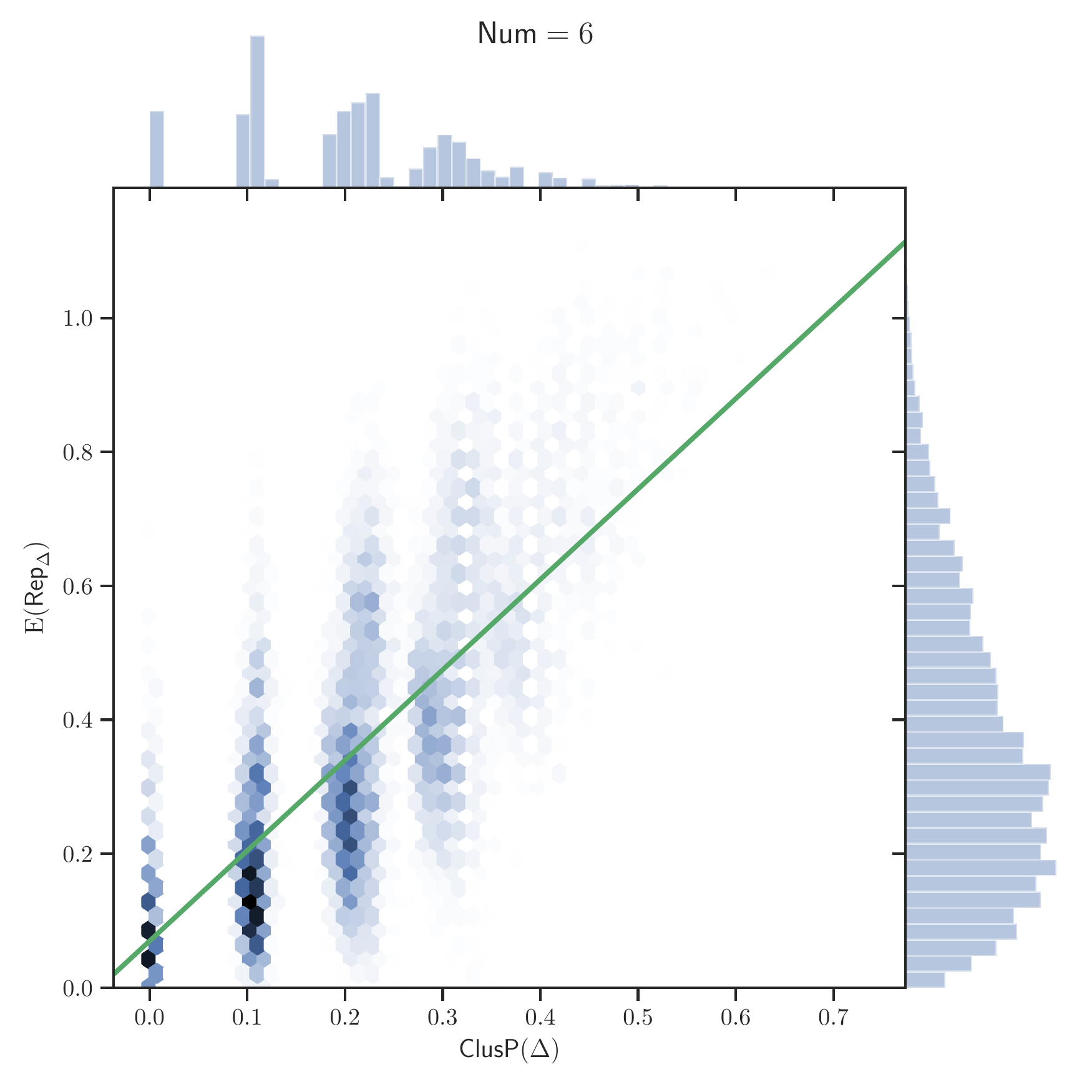}
        \includegraphics[width=5cm]{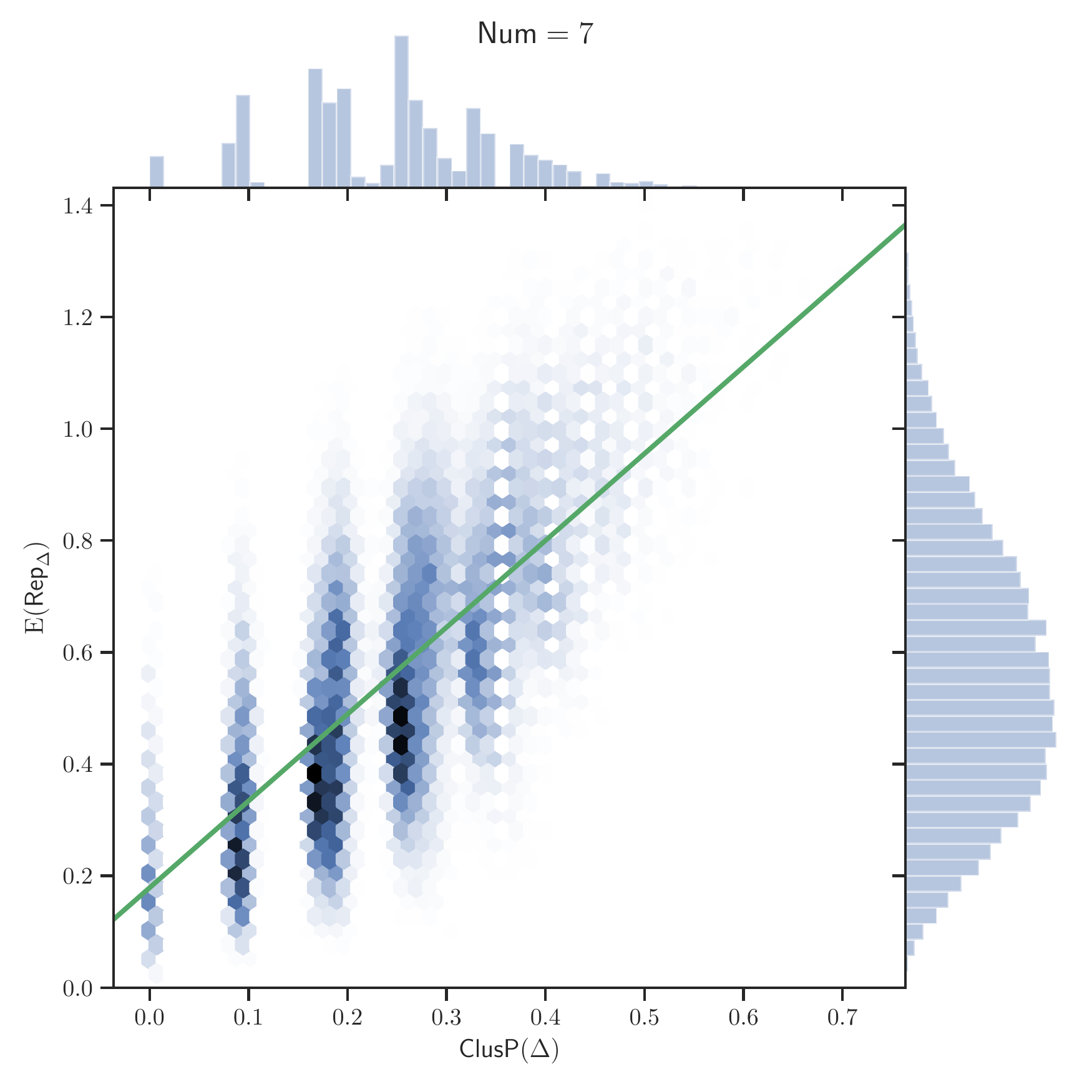}
        \includegraphics[width=5cm]{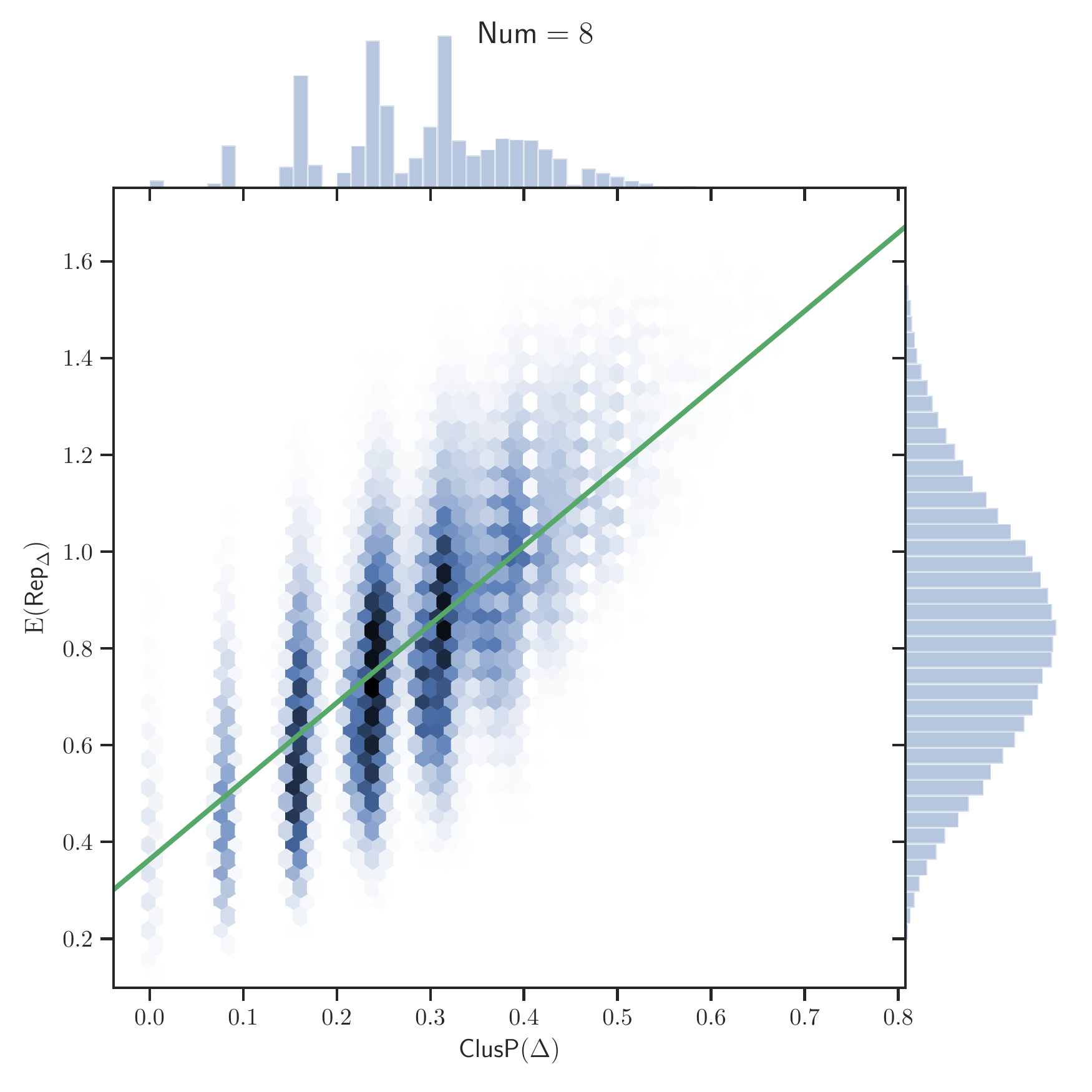}
        \\[1em]
        \includegraphics[width=5cm]{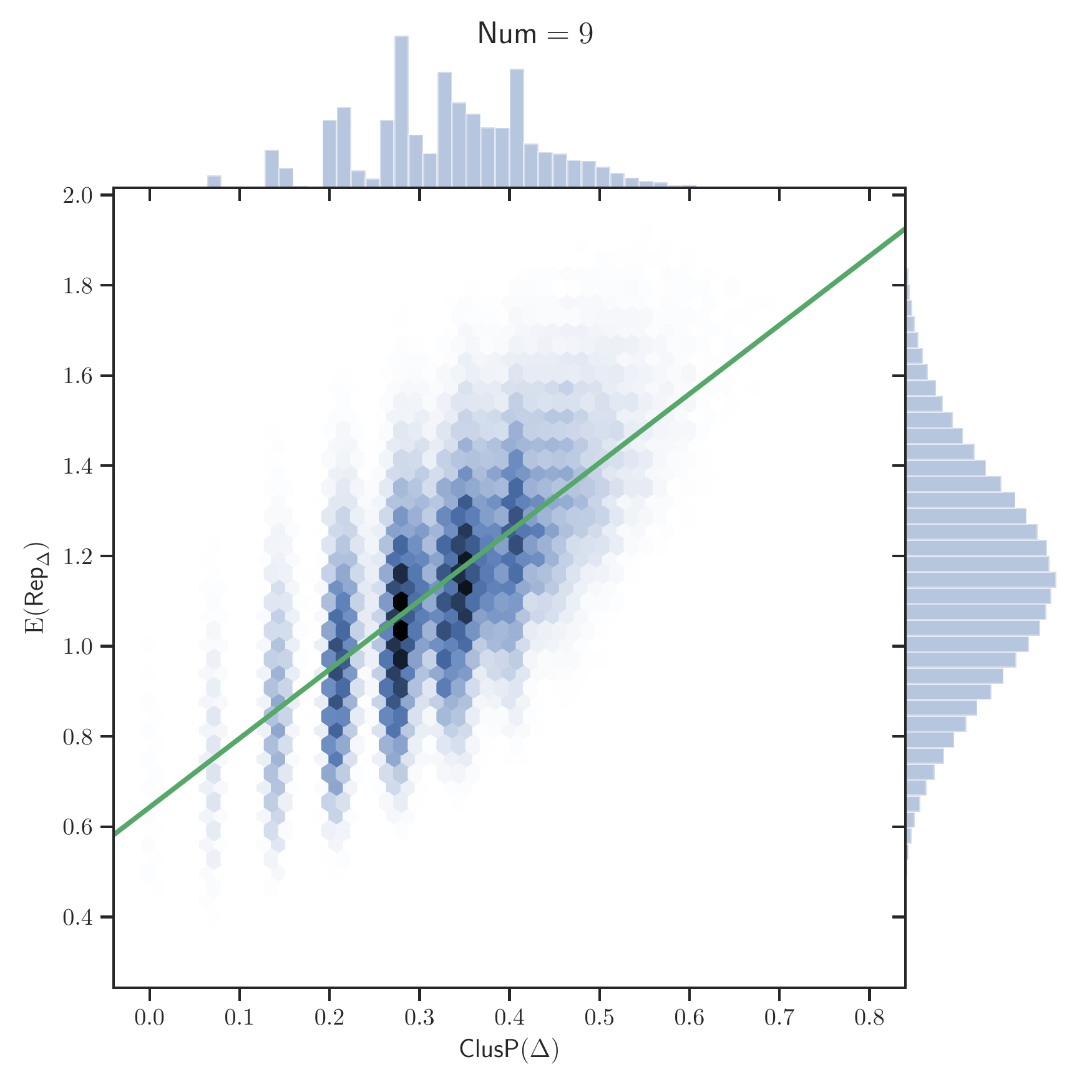}
        \includegraphics[width=5cm]{figures/cluspnumh/ClusP-10-hex.pdf}
        \includegraphics[width=5cm]{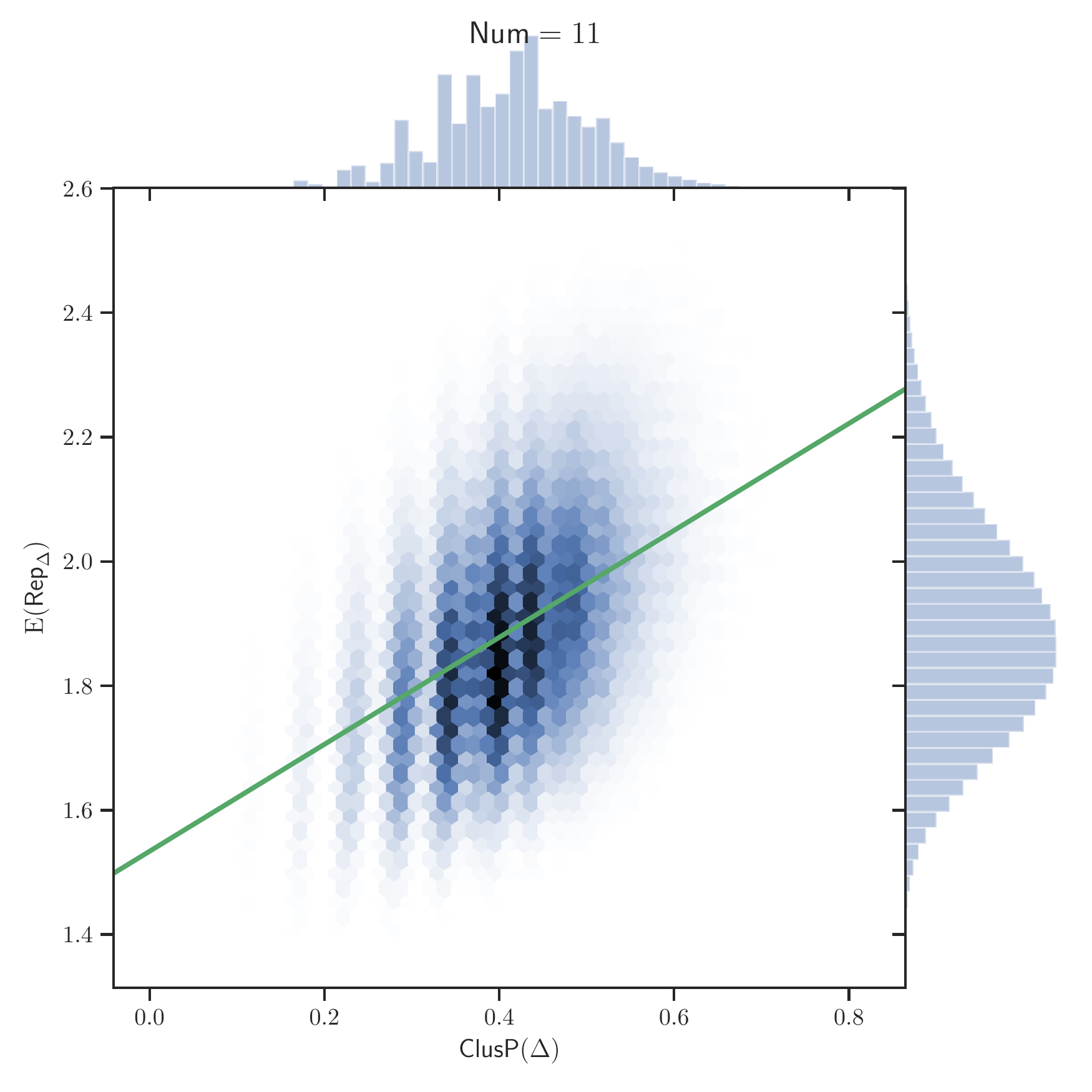}
        \\[1em]
        \includegraphics[width=5cm]{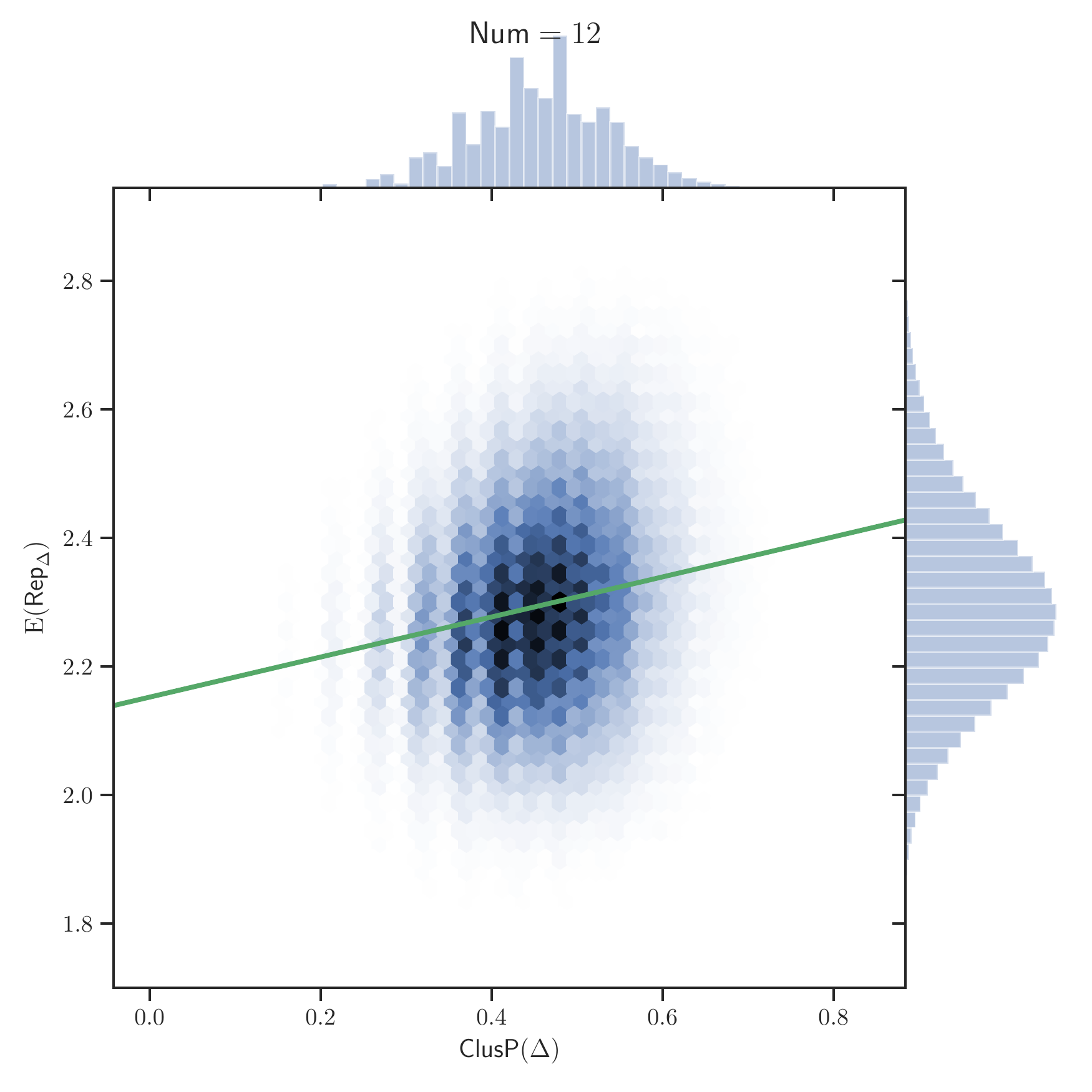}
        \includegraphics[width=5cm]{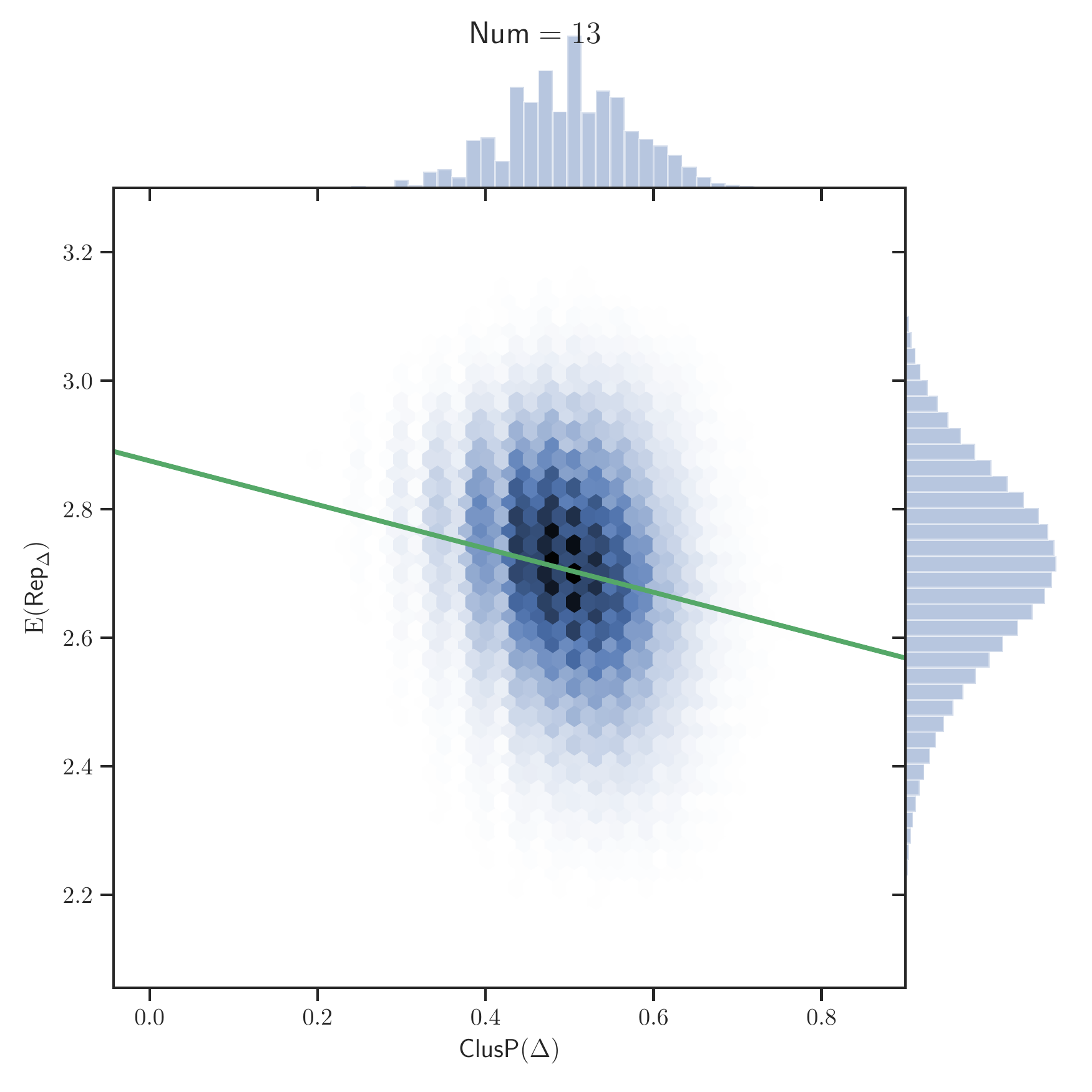}
        \includegraphics[width=5cm]{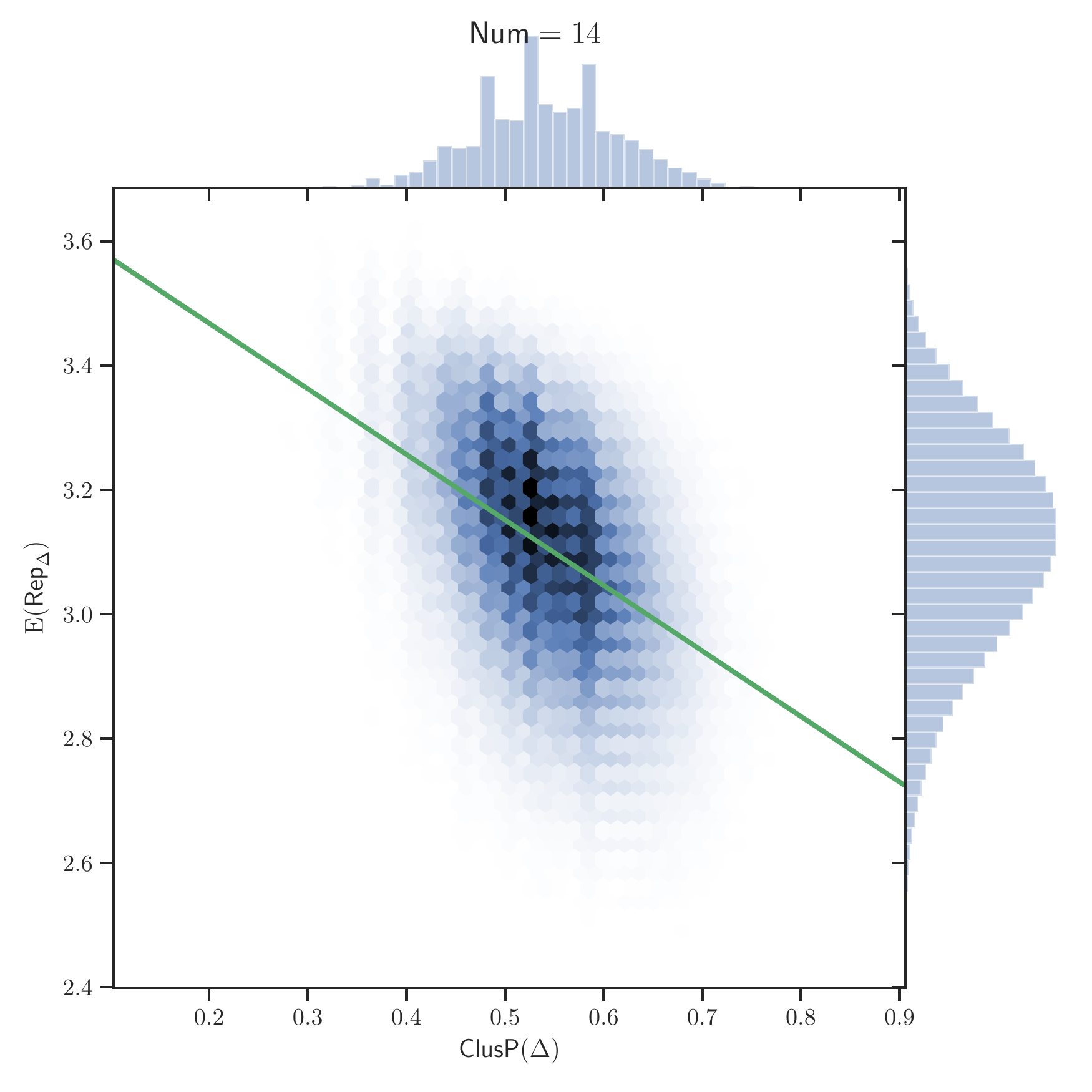}
        \\[1em]
        \includegraphics[width=5cm]{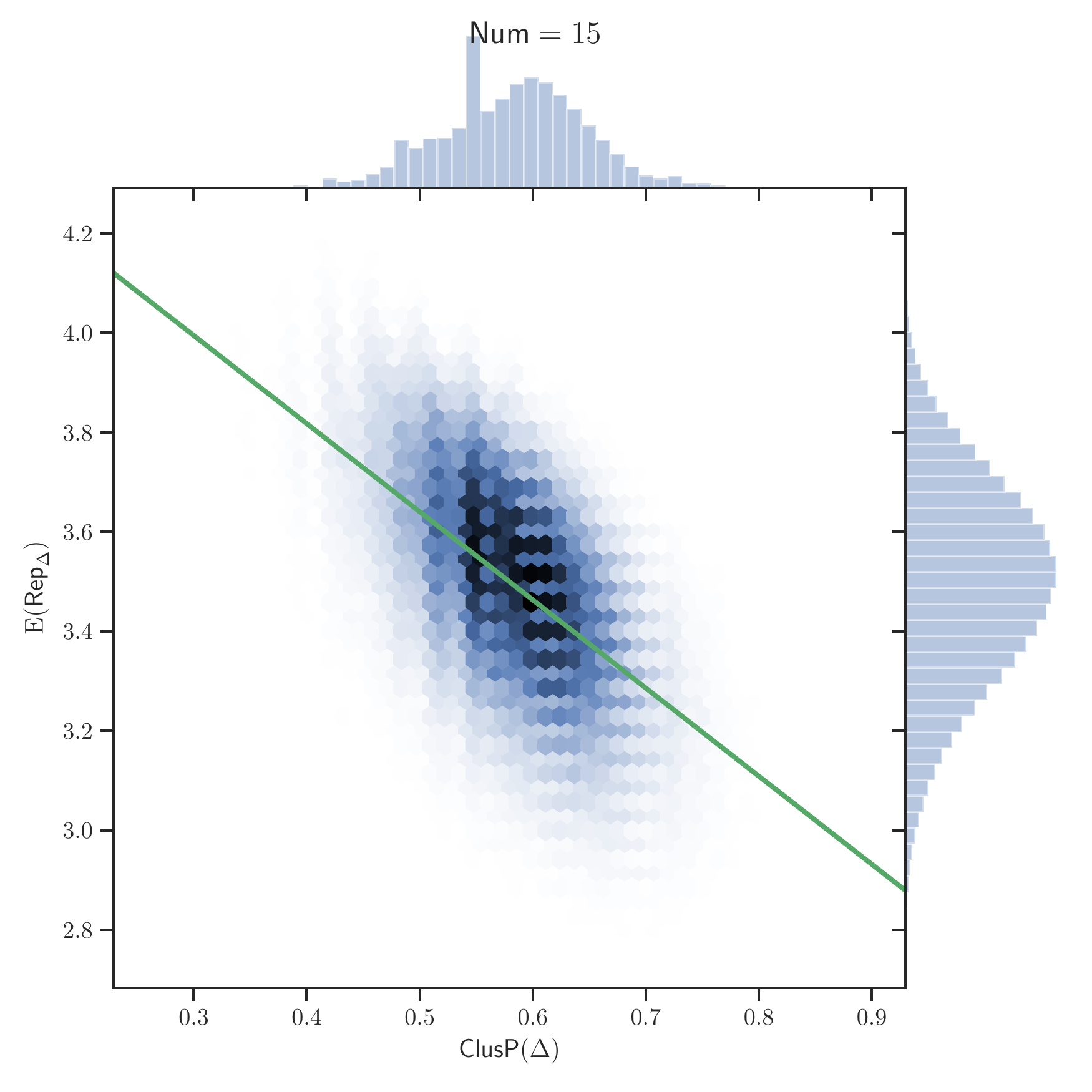}
        \includegraphics[width=5cm]{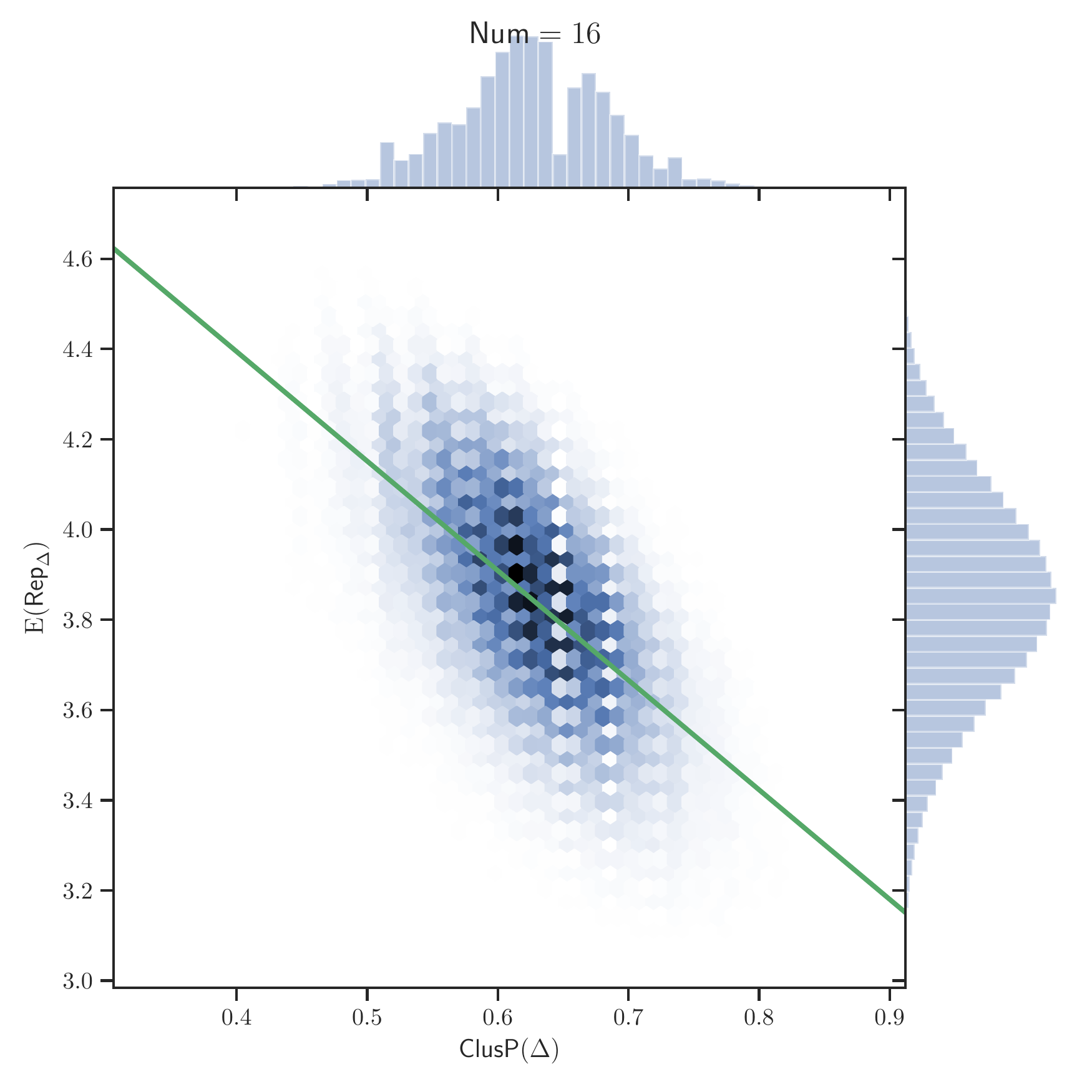}
        \includegraphics[width=5cm]{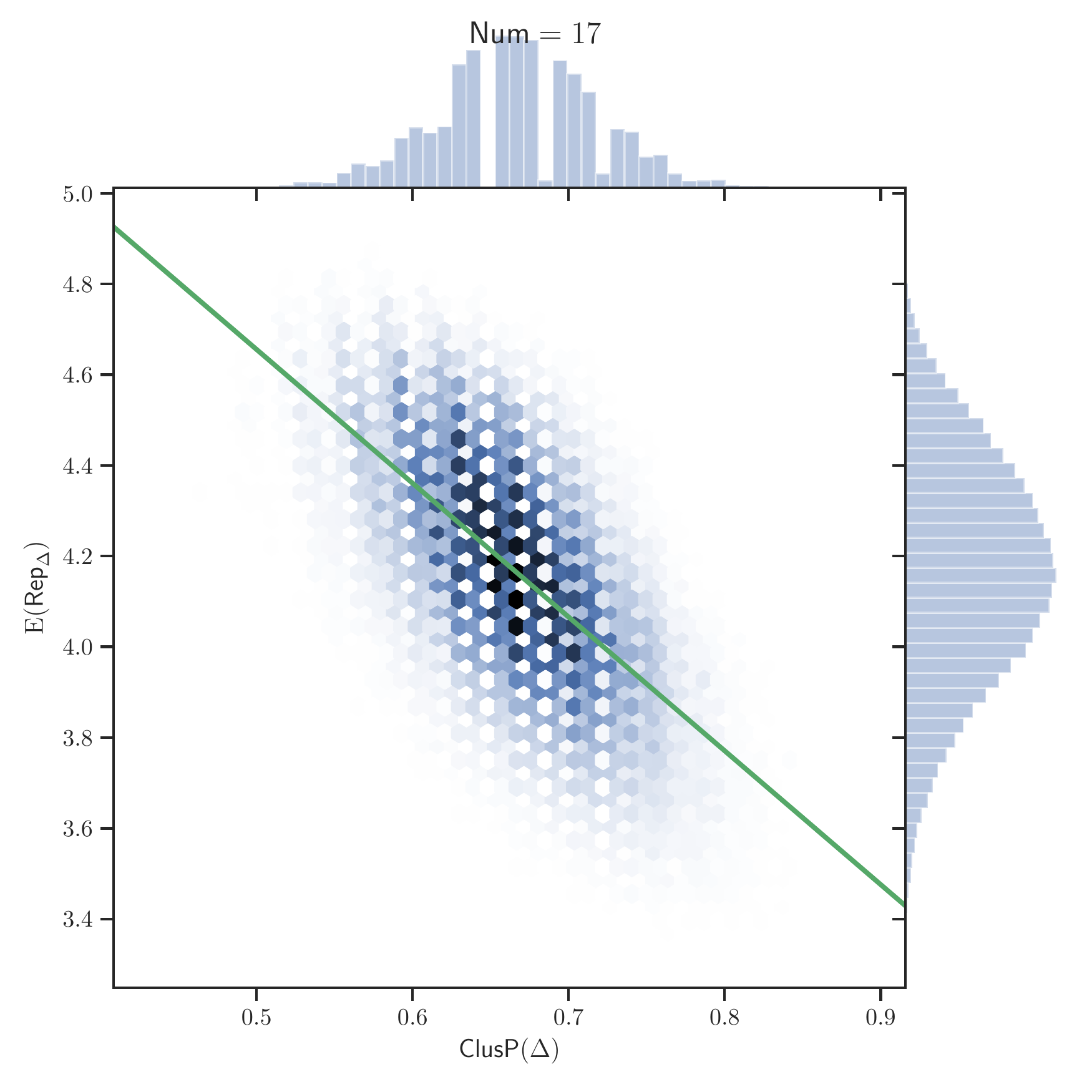}
        \\[1em]
        \includegraphics[width=5cm]{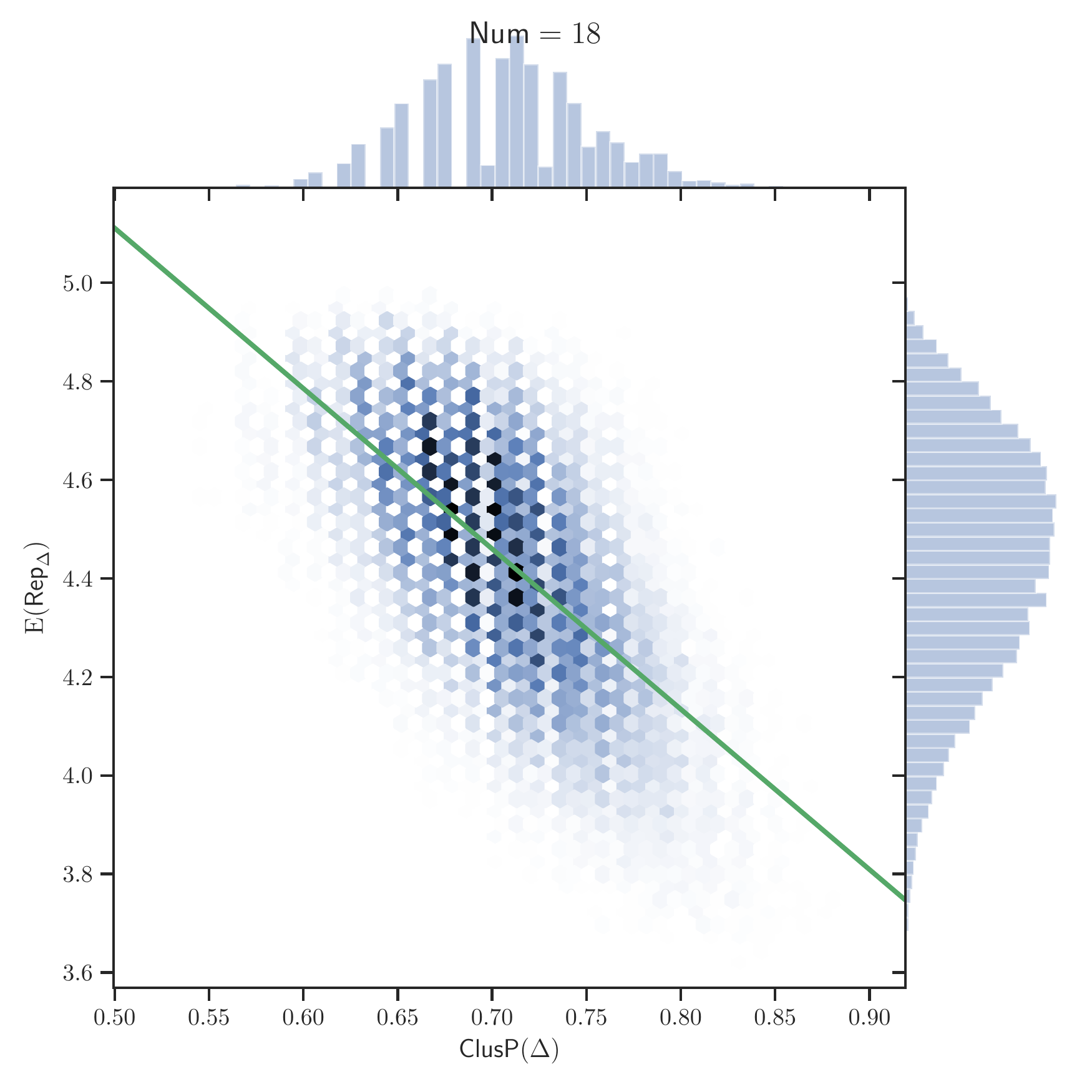}
        \includegraphics[width=5cm]{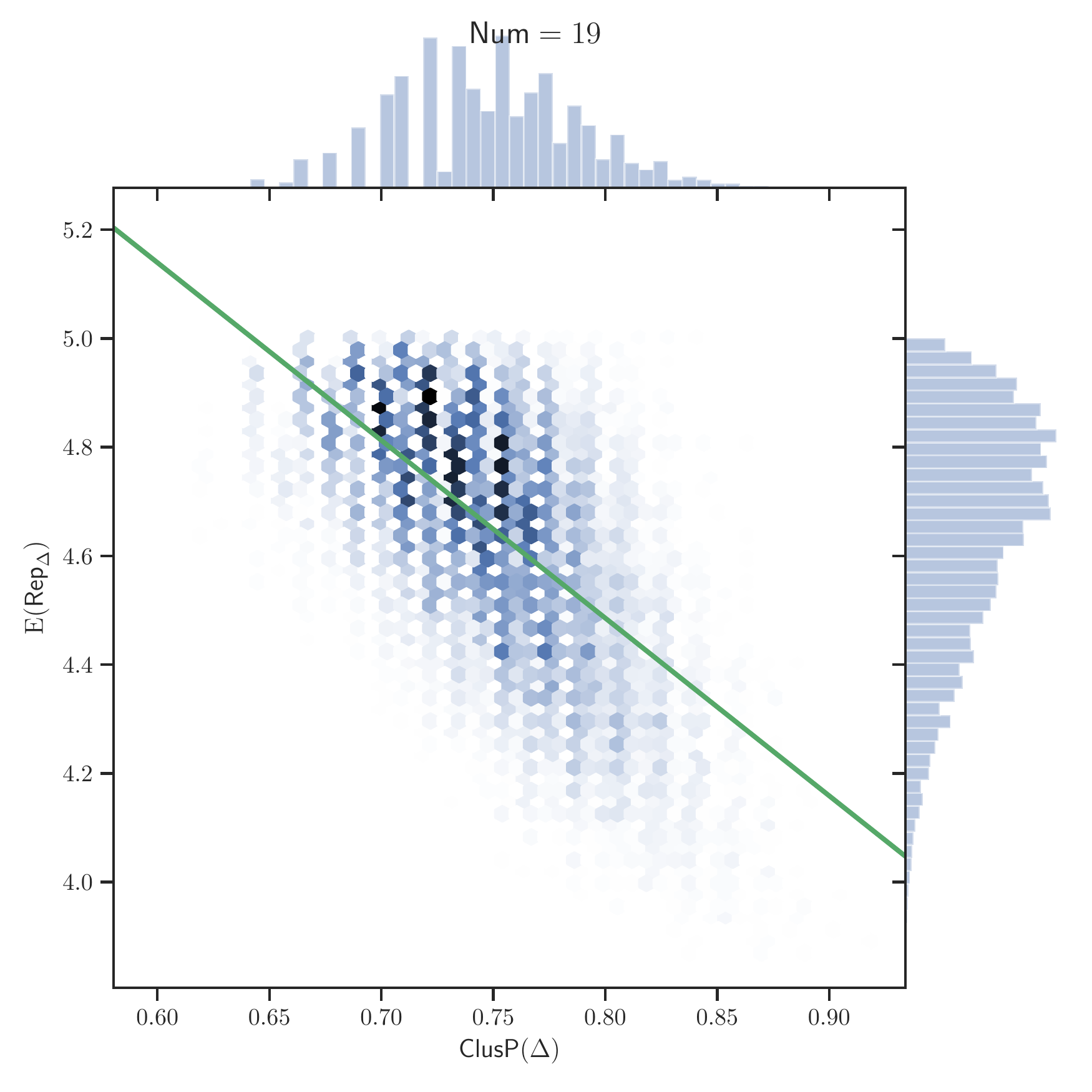}
        \includegraphics[width=5cm]{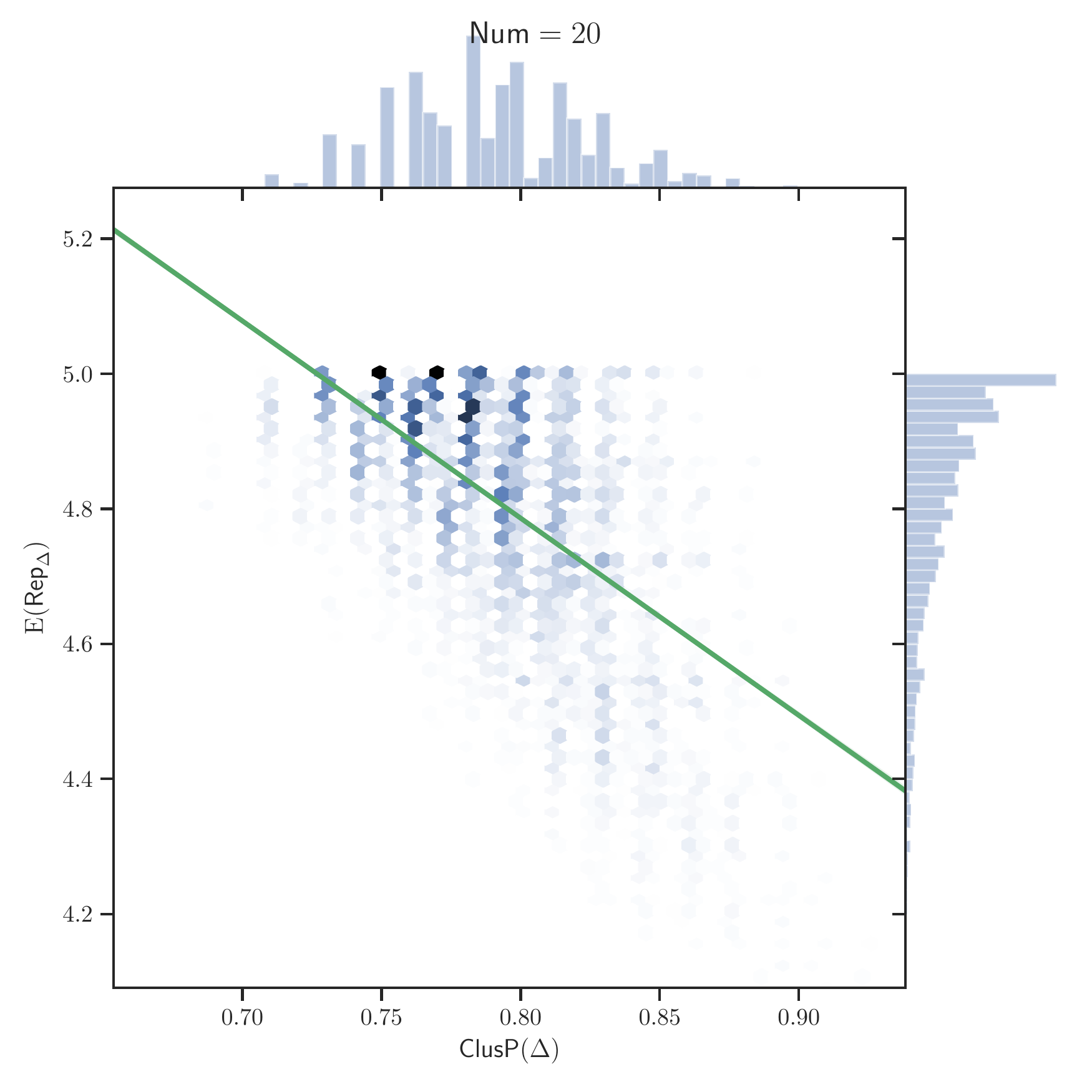}
        \\[1em]
        \includegraphics[width=5cm]{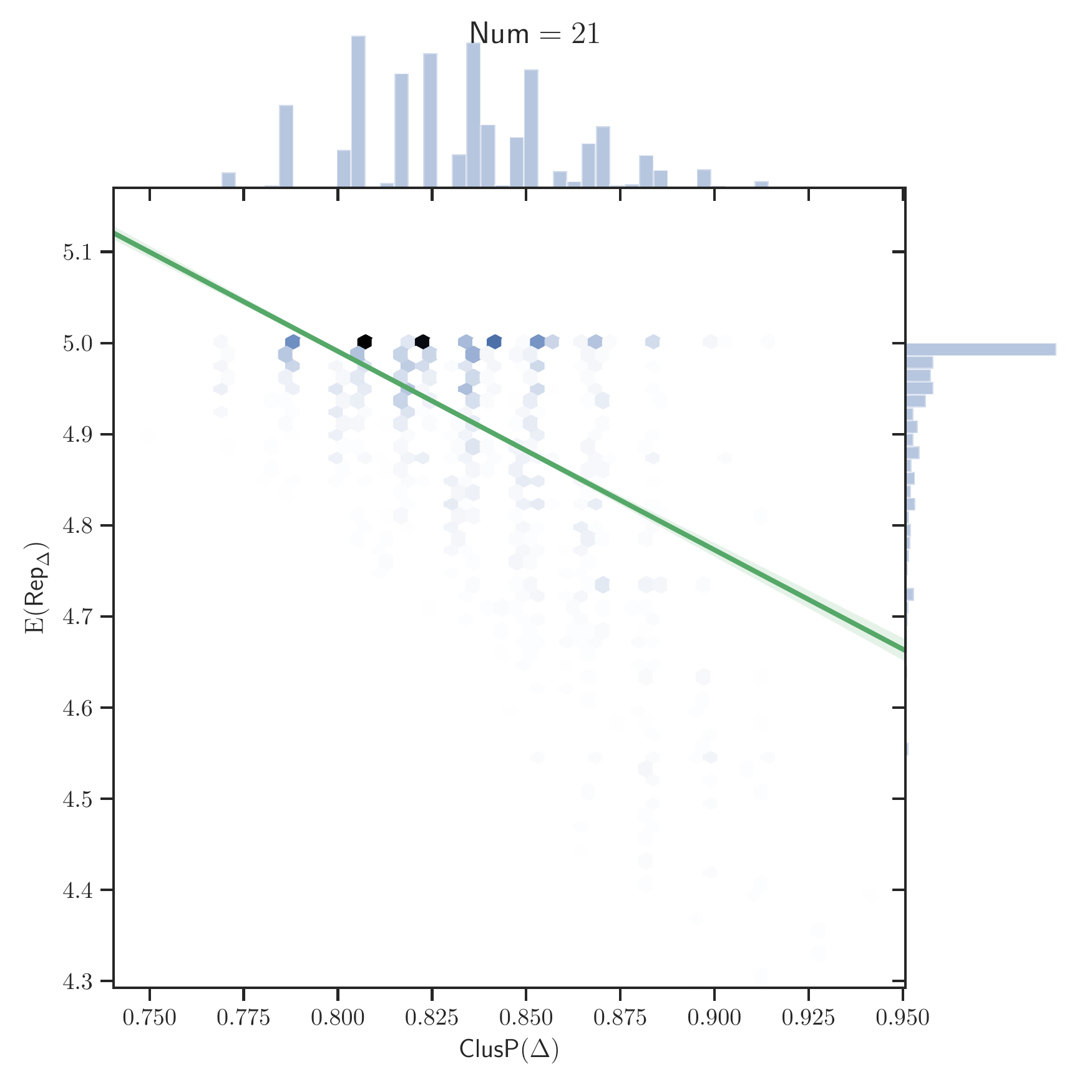}
        \includegraphics[width=5cm]{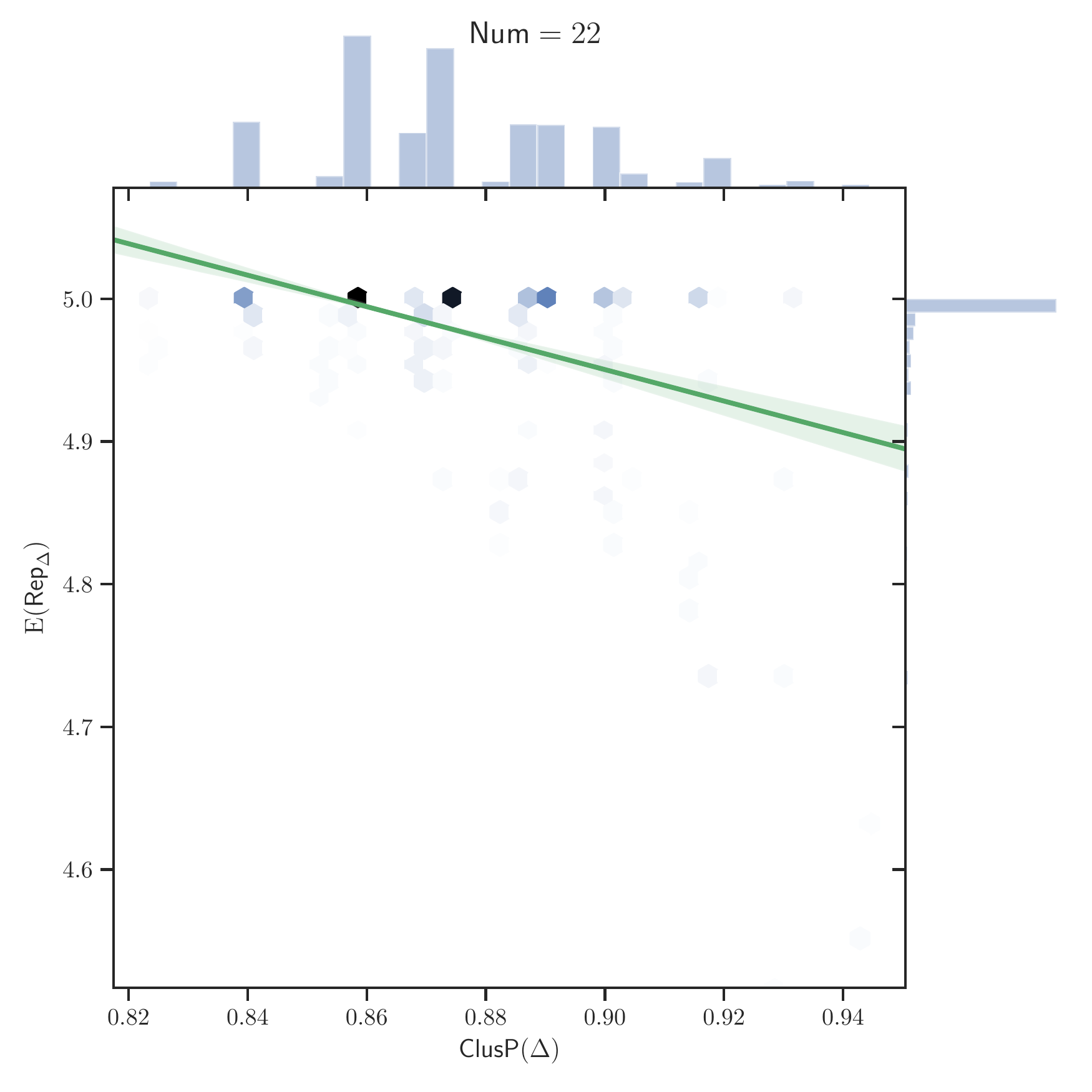}
        \includegraphics[width=5cm]{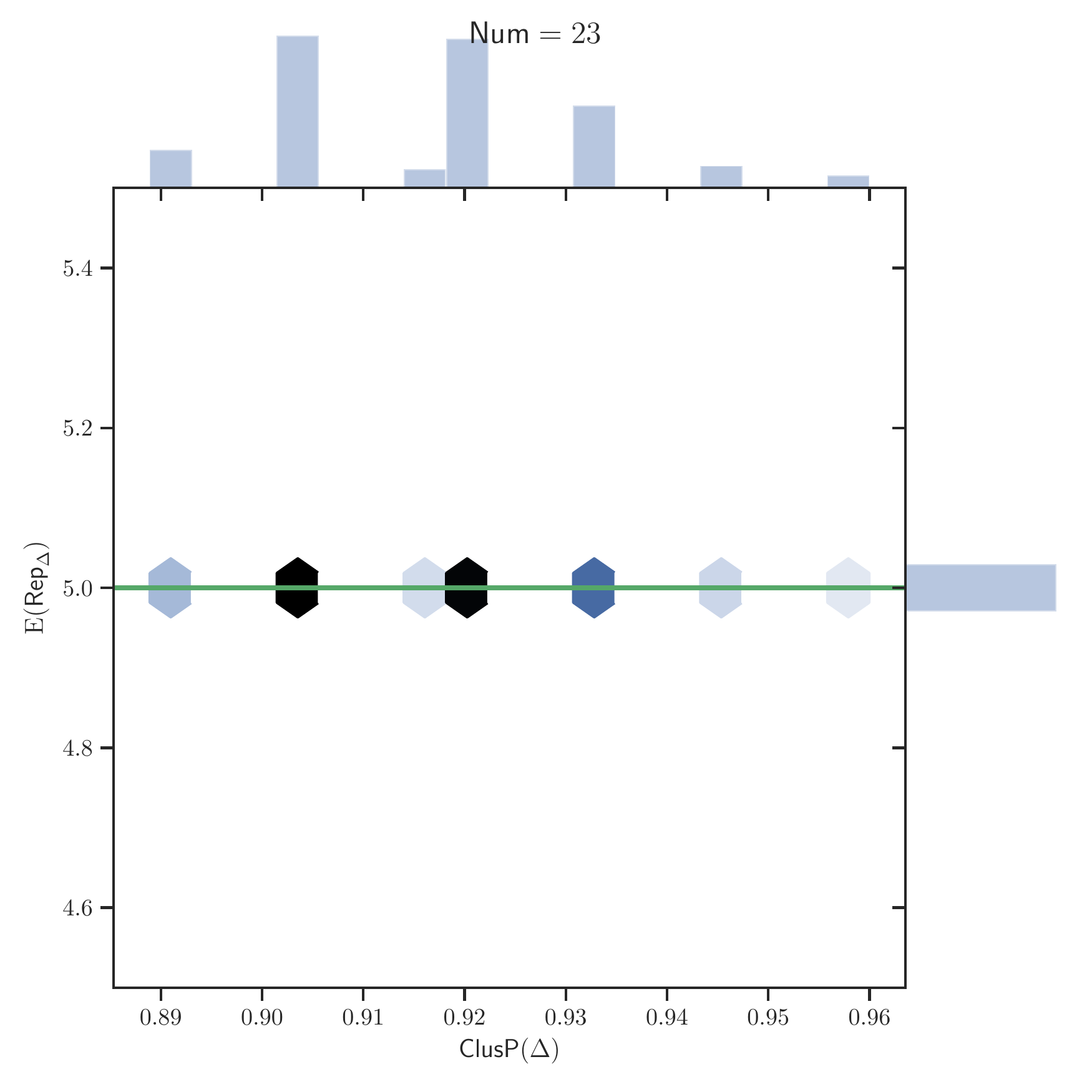}

\begin{landscape}
\section{The rates at which each algorithm approaches the absolute known maximum for a $5\times 5$ grid for differing $k_\mathrm{max}$. ($\mathsf{Num}=5,6,7,8,9,10,11,12$)}
\label{appendix:comparisons}
\includegraphics[scale=0.48, trim={1cm 0.48cm 2cm 1cm},clip]{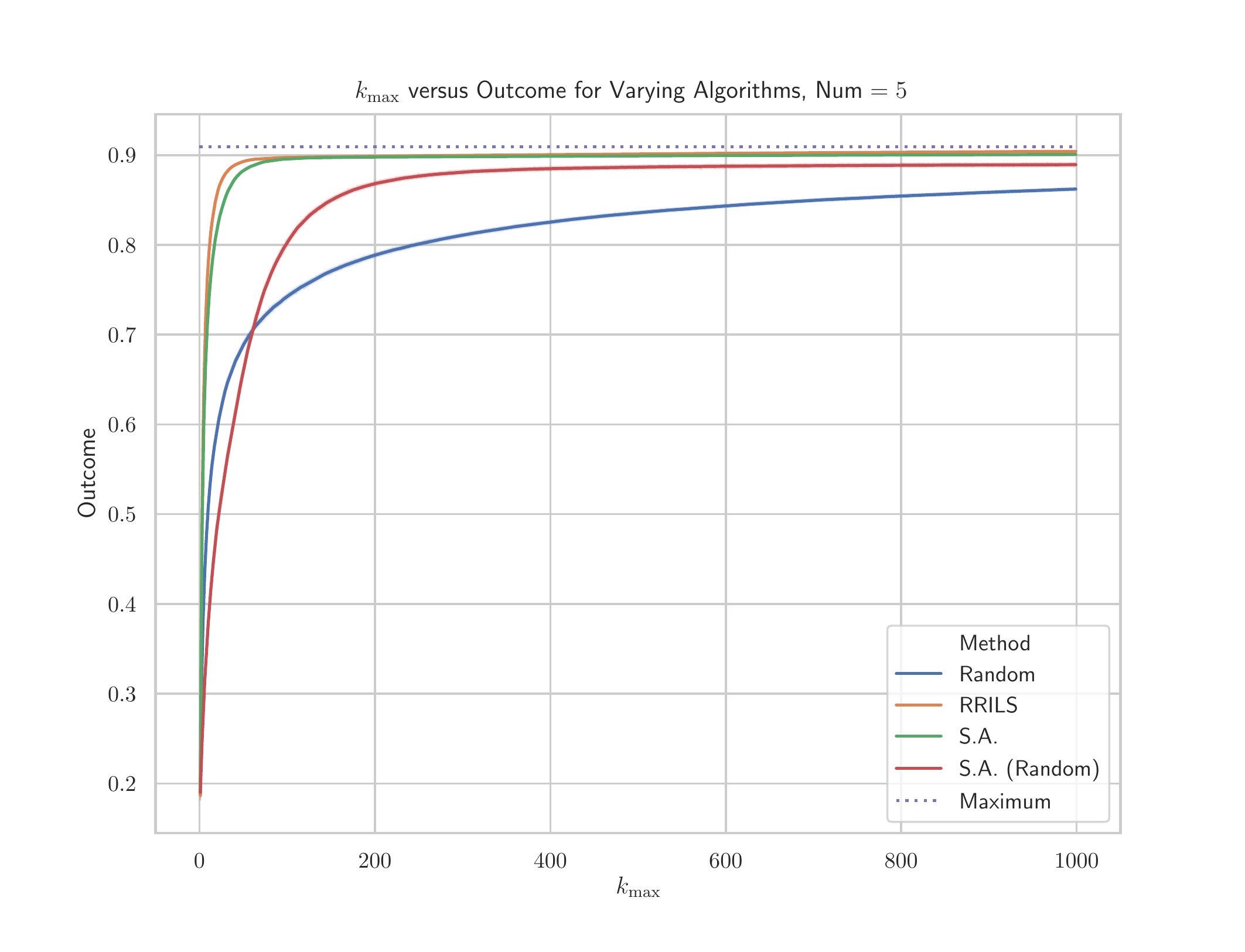}
\includegraphics[scale=0.48, trim={1cm 0.48cm 2cm 1cm},clip]{figures/algorithm-comparison/6.pdf}
\includegraphics[scale=0.48, trim={1cm 0.48cm 2cm 1cm},clip]{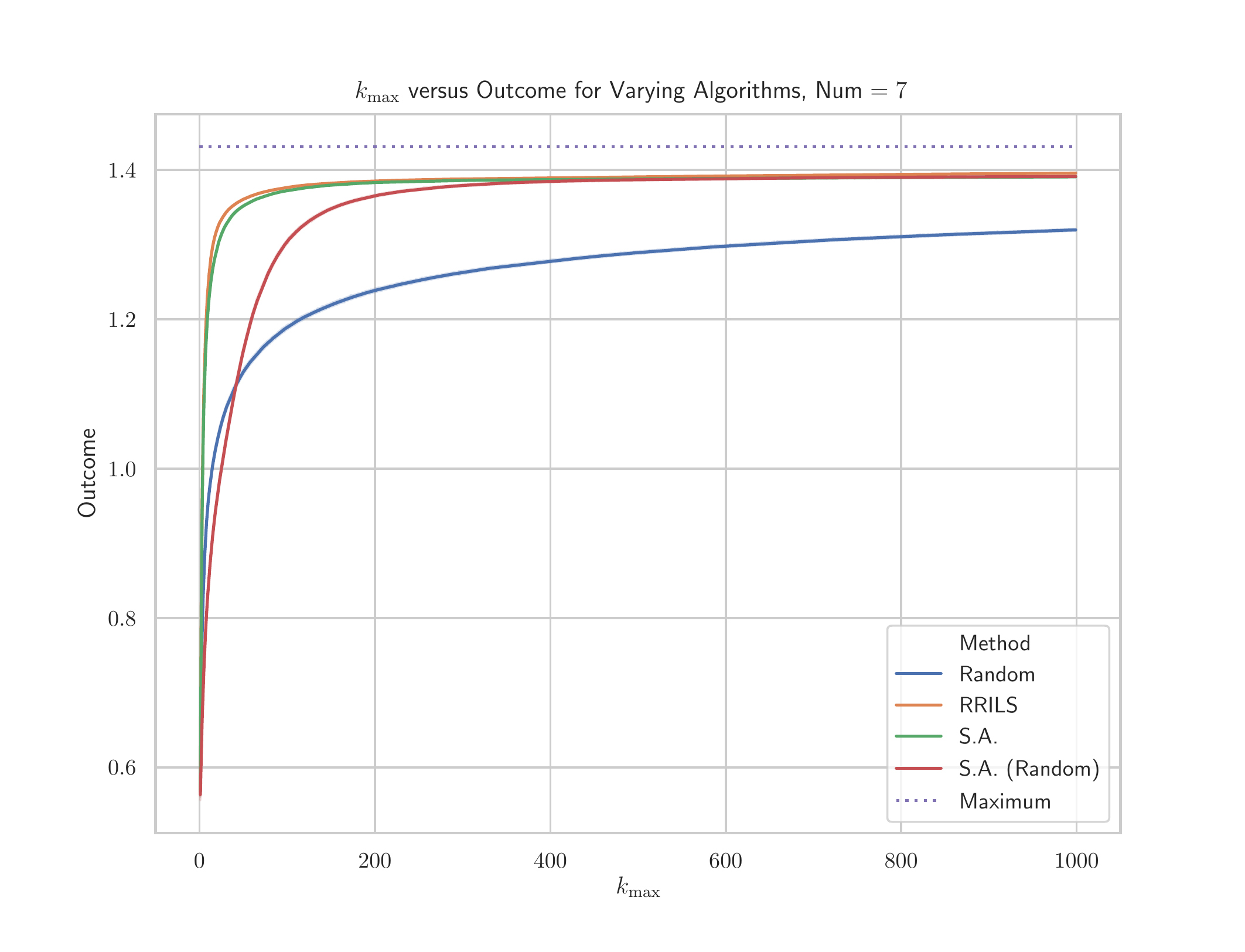}
\includegraphics[scale=0.48, trim={1cm 0.48cm 2cm 1cm},clip]{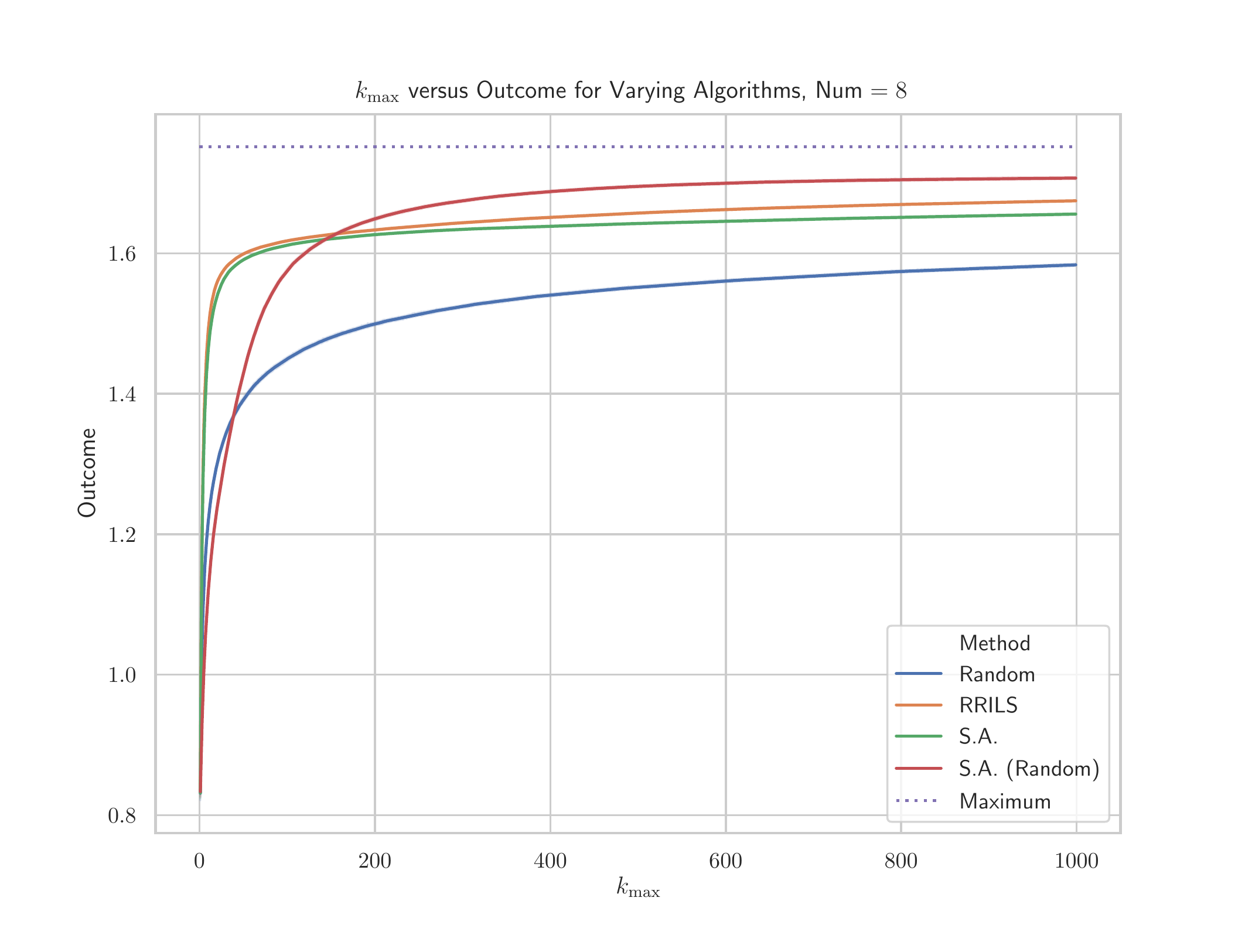}
\includegraphics[scale=0.48, trim={1cm 0.48cm 2cm 1cm},clip]{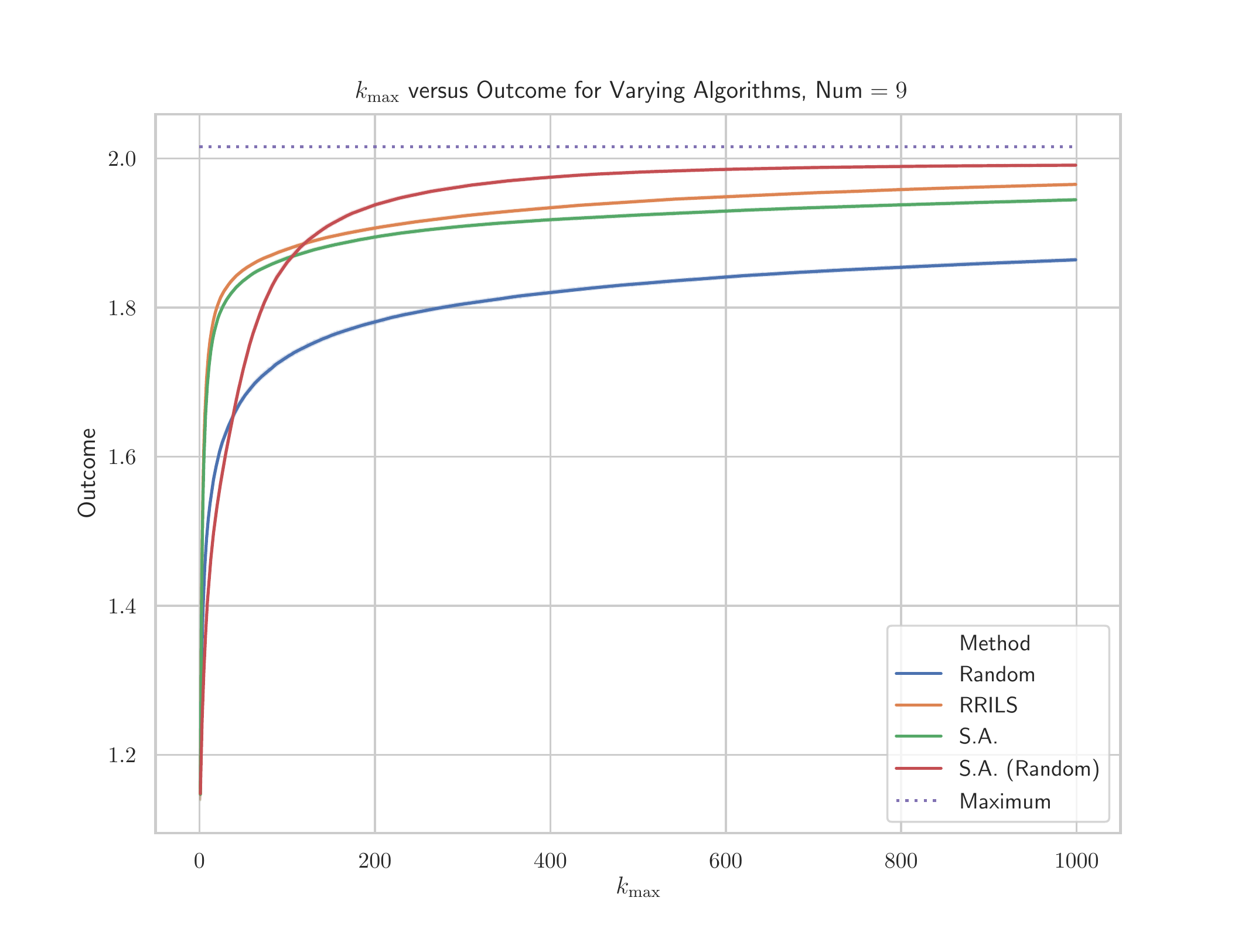}
\includegraphics[scale=0.48, trim={1cm 0.48cm 2cm 1cm},clip]{figures/algorithm-comparison/10.pdf}
\includegraphics[scale=0.48, trim={1cm 0.48cm 2cm 1cm},clip]{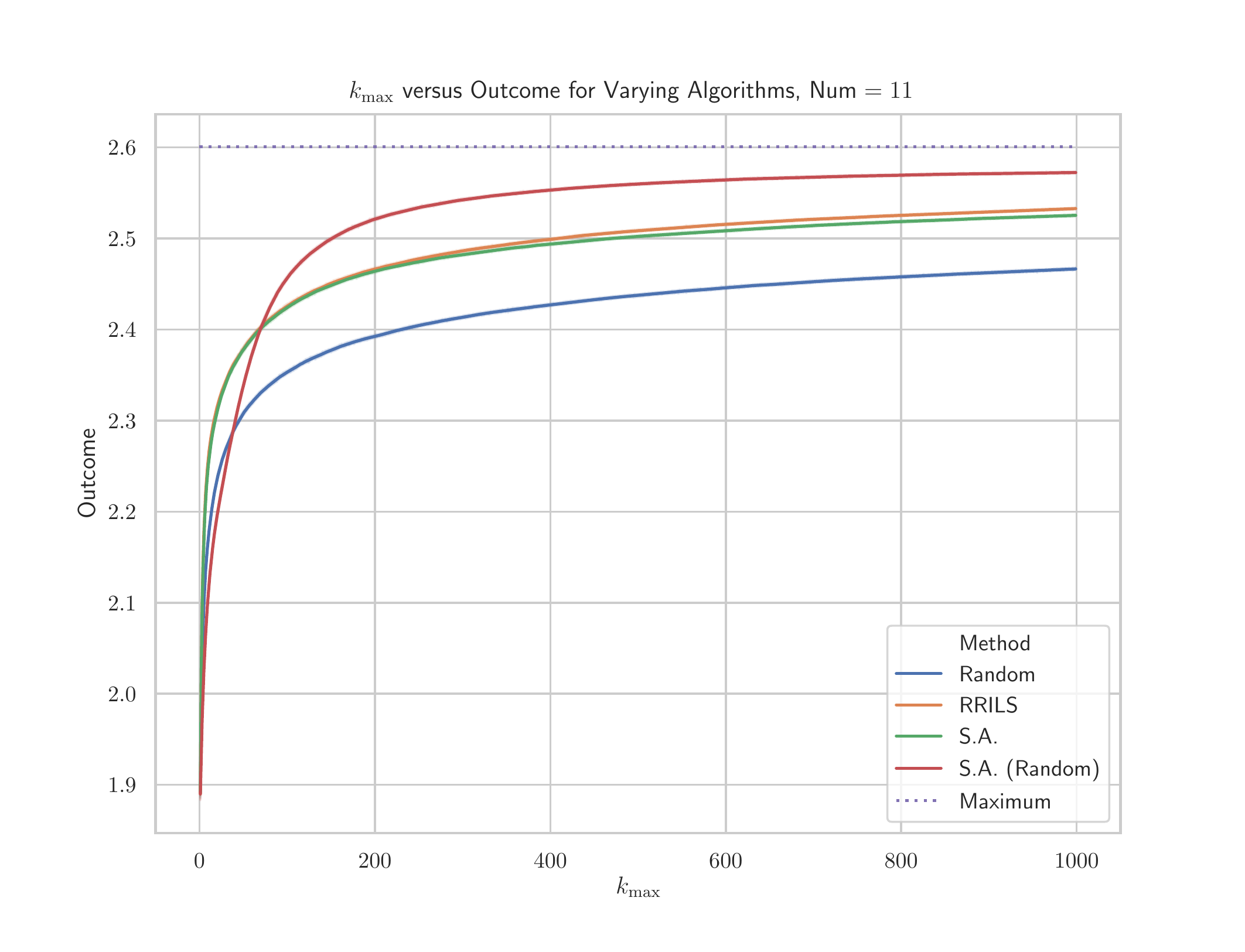}
\includegraphics[scale=0.48, trim={1cm 0.48cm 2cm 1cm},clip]{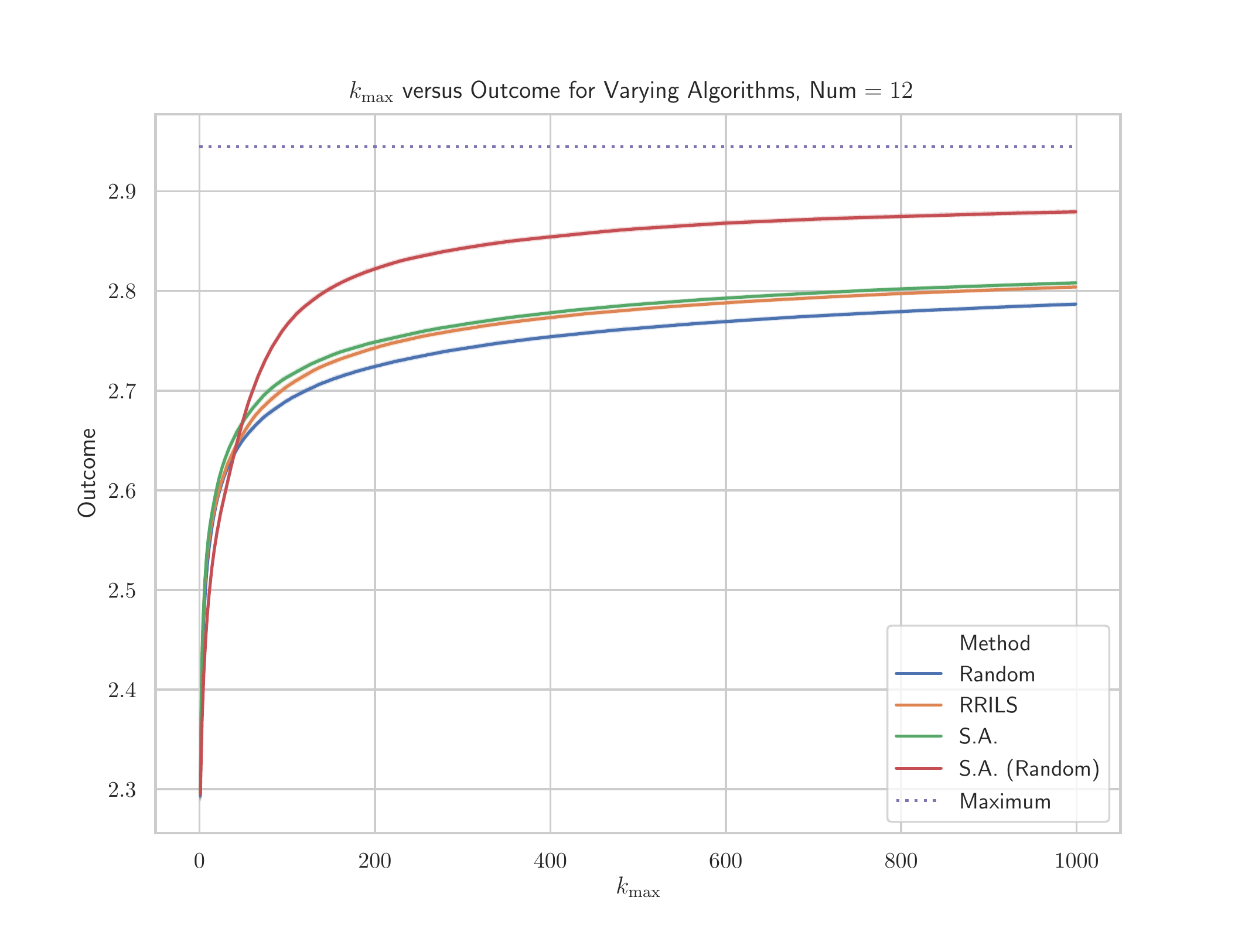}
\end{landscape}

\end{document}